\documentclass[lettersize,journal]{IEEEtran}
\usepackage{amsmath,amsfonts}
\usepackage{algorithmic}
\usepackage{array}
\usepackage{textcomp}
\usepackage{stfloats}
\usepackage{url}
\usepackage{verbatim}
\usepackage{graphicx}
\usepackage{bm}
\usepackage[ruled,linesnumbered]{algorithm2e}
\usepackage{amssymb}
\usepackage{multirow}
\usepackage{caption}
\usepackage{xcolor}
\usepackage{cite} 
\usepackage{mathrsfs}
\usepackage{subfigure}
\usepackage{xurl}

\def\BibTeX{{\rm B\kern-.05em{\sc i\kern-.025em b}\kern-.08em
    T\kern-.1667em\lower.7ex\hbox{E}\kern-.125emX}}
\usepackage{balance}
\usepackage{geometry}
\begin{document}
\title{\huge Reconfigurable Massive MIMO: Precoding Design and Channel Estimation in the Electromagnetic Domain}
\author{Keke Ying, Zhen Gao, Yu Su,~\IEEEmembership{Senior Member,~IEEE}, Tong Qin, Michail Matthaiou,~\IEEEmembership{Fellow,~IEEE}, \\and Robert Schober,~\IEEEmembership{Fellow,~IEEE}

\thanks{The work of Zhen Gao was supported in part by the Natural Science Foundation of China (NSFC) under Grant U2233216 and Grant 62471036; in part by Beijing Natural Science Foundation under Grant L242011; in part by the Shandong Province Natural Science Foundation under Grant ZR2022YQ62; and in part by the Beijing Nova Program. The work of M. Matthaiou was supported by the European Research Council (ERC) under the European Union’s Horizon 2020 Research and Innovation Programme under Agreement 101001331. An earlier version of this paper was presented at the 2023 IEEE Global Communications Conference (GLOBECOM) \cite{GC_ykk} \textit{(Corresponding Authors: Zhen Gao; Tong Qin)} }
		
\thanks{Keke Ying and Tong Qin are with the School of Information and Electronics, Beijing Institute of Technology, Beijing 100081, China (e-mail:ykk@bit.edu.cn; qintong@bit.edu.cn).} %
	
\thanks{Zhen Gao is with State Key Laboratory of CNS/ATM, Beijing Institute of Technology (BIT), Beijing 100081, China, also with BIT Zhuhai 519088, China, also with the MIIT Key Laboratory of Complex-Field Intelligent Sensing, BIT, Beijing 100081, China, also with the Advanced Technology Research Institute of BIT (Jinan), Jinan 250307, China, and also with the Yangtze Delta Region Academy, BIT (Jiaxing), Jiaxing 314019, China (e-mail: gaozhen16@bit.edu.cn).} %

\thanks{Yu Su is with the China Mobile Chengdu Institute of Research and Development, Chengdu 610000, China (e-mail: suyu@cmii.chinamobile.com).}

\thanks{Michail Matthaiou is with the Centre for Wireless Innovation (CWI), Queen’s University Belfast, Belfast BT3 9DT, U.K. (e-mail: m.matthaiou@qub.ac.uk).
} %

\thanks{Robert Schober is with the Institute for Digital Communications, Friedrich-Alexander-University Erlangen-Nürnberg, 91054 Erlangen, Germany (e-mail: robert.schober@fau.de).} %

}


\maketitle

\begin{abstract}	
Reconfigurable massive multiple-input multiple-output (RmMIMO), as an electronically-controlled fluid antenna system, offers increased flexibility for future communication systems by exploiting previously untapped degrees of freedom in the electromagnetic (EM) domain. 
The representation of the traditional spatial domain channel state information (sCSI) limits the insights into the potential of EM domain channel properties, constraining the base station's (BS) utmost capability for precoding design.
This paper leverages the EM domain channel state information (eCSI) for antenna radiation pattern design at the BS. 
We develop an orthogonal decomposition method based on spherical harmonic functions to decompose the radiation pattern into a linear combination of orthogonal bases.
By formulating the radiation pattern design as an optimization problem for the projection coefficients over these bases, we develop a manifold optimization-based method for iterative radiation pattern and digital precoder design. 
To address the eCSI estimation problem, we capitalize on the inherent structure of the channel. Specifically, we propose a subspace-based scheme to reduce the pilot overhead for wideband sCSI estimation.
Given the estimated full-band sCSI, we further employ parameterized methods for angle of arrival estimation. Subsequently, the complete eCSI can be reconstructed after estimating the equivalent channel gain via the least squares method. Simulation results demonstrate that, in comparison to traditional mMIMO systems with fixed antenna radiation patterns, the proposed RmMIMO architecture offers significant throughput gains for multi-user transmission at a low channel estimation overhead. The RmMIMO paves a paradigm shift from conventional  mechanical-controlled fluid antenna systems to an electronically-controlled system, thanks to its pixel-based antennas that adjust the radiation pattern, thereby increasing configuration speed and design flexibility.
\end{abstract}


\section{Introduction}\label{Sec:Intro}
\IEEEPARstart{T}{he} advancement of next-generation wireless communications \cite{Mag_MM,6GMag} is closely tied to the ongoing enhancement of the capabilities of base stations (BSs), where the system capacity predominantly depends on the efficient use of the available physical resources. Numerous studies have focused on optimizing the utilization of the time-frequency resources. In the spatial domain, multiple-input multiple-output (MIMO) technology has evolved into massive MIMO and even extra-large-scale MIMO \cite{XL_MIMO, MM_JSAC_2020}. To decrease the hardware expenditure caused by the large number of radio frequency (RF) chains in fully-digital arrays, phase shifter networks and/or true time delays \cite{MM_JSAC_2021} have been introduced in hybrid analog-digital arrays to provide additional degrees of freedom (DoF).

Nonetheless, in terms of BS deployment, the increase of the antenna aperture poses fundamental challenges, particularly for mast-mounted antennas, which are limited in size by practical windage factors. Additionally, the enlargement of the antenna aperture imposes limitations on its application in scenarios with low loadability, such as unmanned aerial vehicles. Given the increasing demand for enhanced mobile broadband plus (eMBB+) \cite{HuaWei}, there is an urgent necessity to enhance the system capacity for a given limited array aperture. By leveraging the, thus far, unexploited DoF in the electromagnetic (EM) domain \cite{Mag2, BAC_Mag}, reconfigurable antenna architectures present a viable solution for boosting the system capacity.

The EM properties of antennas, encompassing factors such as frequency, polarization, and radiation pattern, remain significantly underutilized today. State-of-the-art metamaterial-based antennas predominantly adjust their amplitude-phase response in the frequency domain via customized antenna structures. Prominent examples include reconfigurable intelligent surfaces (RISs) \cite{RIS}, reconfigurable holographic surfaces (RHSs) \cite{RHS}, and dynamic metasurface antennas (DMAs) \cite{DMA}. Essentially, all these metasurface-based antenna arrays fall into the category of reconfigurable massive multiple-input multiple-output (RmMIMO) systems. These RmMIMO systems provide cost-effective solutions for future BSs by facilitating partial control of the channel state at the EM level. In contrast to the plethora of research works dedicated to exploiting the frequency-domain response properties of reconfigurable antennas, leveraging the polarization \cite{Polar} and radiation pattern \cite{KeMag} has not, to date, received much attention in the literature. This paper focuses on radiation pattern (RP)-reconfigurable massive MIMO systems. For convenience, we still refer to these systems simply as RmMIMO in the remainder of the paper.

\subsection{Related Works}
RmMIMO can be implemented based on various hardware platforms \cite{Mag1, Mag2, Hardware}. The origins of RP-reconfigurable antennas can be traced back to the reactively controlled directive arrays introduced in the 1970s \cite{Harrington}. The fundamental concept here involves leveraging the mutual coupling between an active antenna (fed by a single RF chain) and multiple parasitic antennas. By altering the reactive loading of these parasitic elements, the direction of radiation pattern can be adjusted. This concept is also known as an electronically steerable passive array radiator (ESPAR) antenna \cite{Japan, HKUST_ESPAR}. With advances in hardware fabrication methodologies, more compact antenna structures have been devised.  Pixel antennas \cite{RPA, RPA2, HKUST_RPA, HKUST_360, RPA_BF}, which can also be viewed as a kind of electronically-controlled fluid antenna \cite{KKW-1, KKW-2}, offer a reconfigurable aperture that divides the radiating elements into a large number of electrically small parasitic elements, referred to as pixels. These pixels enable changes in the antenna's radiation pattern using cost-effective PIN diodes. For instance, in a notable example \cite{HKUST_360}, a highly RP-reconfigurable planar antenna was designed with an Alford loop and electronically tunable compact parasitic pixel rings to provide reconfigurability, achieving $360^{\circ}$ single- and multi-beam steering. Moreover, reconfigurable pixel antennas can not only alter the radiation pattern direction but also shape the pattern itself. With the addition of an extra parasitic layer above the patch layer, a compact hardware structure was developed in \cite{RPA_BF}, capable of generating different radiation pattern shapes by simply adjusting the PIN diode connections. 

RmMIMO systems \cite{DoF1, DoF2, DoF3, DoF4} can offer additional DoF for manipulating antennas at the EM level. In \cite{DoF1, DoF2}, the Gram-Schmidt orthonormalization procedure was introduced in an ESPAR antenna system to determine the number of orthogonal basis patterns that support uncorrelated signals in the beamspace domain. To overcome the high computational complexity of the Gram-Schmidt method, a characteristic modes method \cite{DoF3} and a pattern correlation decomposition method \cite{DoF4} were proposed, which are applicable to ESPAR antennas in single-RF MIMO systems.

The introduction of additional DoF for radiation pattern design also requires corresponding optimization methods to customize the radiation patterns for alignment with the channel environment. Based on the available radiation pattern options, existing pattern designs employ discrete space pattern selection or continuous space pattern optimization. For discrete space pattern selection, the task is to find suitable combinations of predefined radiation patterns in mMIMO systems to maximize the overall system performance. This involves developing effective search strategies within a huge pattern search space. For instance, in a multiple-input single-output (MISO) system with 16 antennas, each having four candidate radiation patterns, the search space encompasses a total of $4^{16}$ possibilities, which is prohibitively large for exhaustive exploration.
In \cite{RPA_BF}, an iterative mode search (IMS) method was devised to optimize the radiation pattern combinations at the transmitter. While IMS focuses on pattern optimization within a limited set, its complexity scales linearly with the number of transmit antennas. Similar methods have been proposed in \cite{KeMag, HW_Shanghai}, underscoring the substantial potential of improving the system throughput by employing reconfigurable patterns in MIMO systems. Furthermore, in \cite{TS}, multi-armed bandit-based online learning algorithms were developed, which exploit the channel correlations for different radiation pattern states to accelerate pattern selection policy convergence. However, the small-scale single-user scenario considered in \cite{TS} may not be directly extendable to mMIMO systems with multi-user data transmission.
For continuous space pattern optimization, the authors of \cite{XJ_Pre_SU, XJ_Pre_MU} formulated the pattern optimization as a continuous space sampling matrix design problem. In \cite{XJ_Pre_SU}, a sequential optimization framework was proposed to enhance the pattern design in the continuous pattern space for MIMO arrays. Regarding the multi-user downlink precoding problem, the joint optimization of symbol-level precoding in the digital domain and pattern design in the EM domain was considered in \cite{XJ_Pre_MU}.

In \cite{MRA_CE, XJ_CE}, the authors tackled the channel estimation challenges pertaining to RmMIMO systems. Unlike conventional mMIMO systems that typically involve the estimation of a single channel matrix, RmMIMO systems, with their different radiation patterns, may feature multiple different channel states, leading to substantial pilot overhead.
In \cite{MRA_CE}, using the Gram-Schmidt procedure, the radiation patterns were decomposed into orthogonal basic patterns, effectively decoupling the antenna radiation patterns from the remainder of the channel environment. This separation facilitated the development of a joint channel estimation and prediction scheme, ultimately mitigating the high overhead associated with channel estimation for different antenna radiation patterns. 
On the other hand, the authors of \cite{XJ_CE} introduced a deep learning-based channel extrapolation method. During the channel estimation stage, different patterns were assigned to different antennas at the transmitter, and a deep neural network was employed at the receiver to extrapolate the channels for the other patterns. This approach yields good channel estimation performance at low pilot overhead.

\subsection{Our Contributions}
Despite the above progress in the RmMIMO system design, there are persisting challenges that still need to be solved. Existing works \cite{KeMag, RPA_BF, HW_Shanghai,TS} focused on discrete-space pattern selection, which cannot fully utilize the DoF in the radiation pattern space. For the continuous-space radiation pattern optimization problem, only single-user transmission was considered in \cite{XJ_Pre_SU}. For the multi-user narrowband transmission design in \cite{XJ_Pre_MU}, the same radiation pattern was assumed for all BS antennas. However, the efficient design of multi-user wideband RmMIMO systems, where different antennas can adopt different radiation patterns, remains an open problem. On the other hand, the channel estimation methods in \cite{MRA_CE, XJ_CE} can only estimate or predict channels for patterns within a limited discrete set. When channels for the design of arbitrary radiation patterns in the continuous space have to be estimated, these methods cannot be employed. 

Against the above background, we aim to establish a new framework that solves the EM domain precoding and channel estimation problems for continuous radiation pattern design. The main contributions of this paper can be summarized as follows:

$\bullet$ We introduce a novel spherical harmonious functions-based orthogonal decomposition method for radiation patterns. This technique simplifies the formulation of the pattern optimization and channel estimation problems in the EM domain for RmMIMO systems. Unlike traditional spatial domain channel state information (sCSI), we introduce the concept of EM domain channel state information (eCSI) to disentangle the impact of the radiation patterns from other aspects of the channel environment. The eCSI provides a more detailed description of the channel characteristics in the EM domain, which also provides essential information for continuous-space radiation pattern design.

$\bullet$ We propose an alternating optimization method for radiation pattern design and digital domain precoding for wideband multi-user downlink transmission. Specifically, we transform the pattern design problem into a projection coefficient optimization problem over orthogonal bases consisting of spherical harmonious functions. We then employ manifold optimization to iteratively update the radiation patterns and digital precoders. The proposed approach encompasses the cases where all antennas at the BS use either the same pattern or different patterns.

$\bullet$ We propose a cost-effective wideband eCSI estimation method comparable in pilot overhead to channel estimation methods for traditional massive MIMO (TmMIMO) systems. In particular, we first estimate the multipath delay based on a one-dimensional ESPRIT\footnote{Here, ESPRIT is short for Estimation of Signal Parameters via a Rotational Invariance Technique \cite{Liao_Tcom}.} (1D-ESPRIT) algorithm, which aims to recover the full-band sCSI from undersampled frequency domain observations. Leveraging the estimated sCSI, we introduce a two-dimensional ESPRIT (2D-ESPRIT) algorithm, thereby facilitating a parameterized angle of arrival (AoA) estimation. Then, affording more pilot symbols than the number of multipaths, the equivalent channel gain of the eCSI is acquired by a simple least squares (LS) algorithm. Finally, based on the proposed framework, the eCSI is effectively reconstructed and used to regenerate the sCSI for arbitrary customized radiation patterns.

$\bullet$ We conduct extensive simulations to evaluate the performance of the proposed schemes. Our simulation results show that a pilot overhead comparable to that of TmMIMO systems is sufficient to ensure good eCSI estimation performance, while significant spectral efficiency (SE) improvements can be obtained with the proposed RmMIMO system design compared to TmMIMO.

The remainder of this paper is organized as follows: Section~\ref{S2} introduces the RmMIMO system model and the problem formulation for radiation pattern optimization. Section~\ref{S3} investigates the EM domain radiation pattern and digital domain precoder design, where the cases of adopting the same/different radiation patterns for all BS antennas are considered. The sCSI and eCSI estimation problems are tackled in Section~\ref{S4}. Simulation results are presented in Section~\ref{S6}. Finally, we draw our conclusions in Section~\ref{S7}.

\textit{Notation}: Matrices and column vectors are denoted by uppercase and lowercase boldface letters, respectively; $(\cdot)^{\rm T}$, $(\cdot)^{*}$, $(\cdot)^{\rm H}$, $(\cdot)^{-1}$, and $\mathbb{E}\left\{\cdot\right\}$ denote the transpose, conjugate, Hermitian transpose, inversion, and expectation operations, respectively. The $(i,j)$-th entry of matrix $\bm{A}$ is denoted as $\left[\bm{A}\right]_{i,j}$; $\text{diag}\left(\bm{a}\right)$ transforms the vector $\bm{a}$ into the corresponding diagonal matrix; $\text{Blkdiag}\{\bm{a}_{1},\bm{a}_{2},\ldots, \bm{a}_{M}\}$ operation constructs a block diagonal matrix by arranging vectors $\{\bm{a}_i\}_{i=1}^{M}$ along the diagonal. $\|\bm{A}\|_{\rm F}$ denotes the Frobenius norm of $\bm{A}$; $\mathcal{R}e\left(\bm{A}\right)$ refers to the real part of matrix $\bm{A}$, $\arg\left(a\right)$ represents the phase of complex number $a$. Besides, $\bm{1}_{N}$ denotes the vector of size $N$ with all the elements being $1$;
$\otimes$, $\circledast$, and $\odot$ denote the Kronecker, Khatri-Rao, and Hadamard products, respectively; $\lfloor a \rfloor$ represents the floor function operation. Finally, $\mathcal{U}[a,b]$ refers to a uniform distribution over the interval $[a,b]$ while $\mathcal{N}_{c}\left(\mu, \sigma^2\right)$ denotes a complex Gaussian distribution with mean $\mu$ and variance $\sigma^2$.

\section{System Model and Problem Formulation}\label{S2}
\subsection{Downlink Transmission Model}\label{S3.2}
\begin{figure}[!t]
	\begin{center}
		\includegraphics[width = 1\columnwidth]{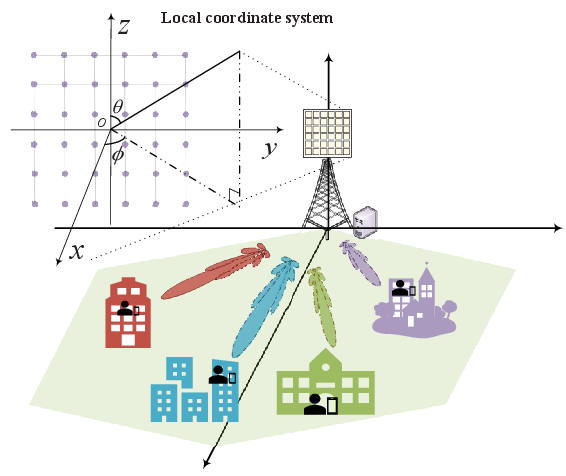}
	\end{center}
	\vspace*{-3mm}
	\captionsetup{font={footnotesize}, singlelinecheck = off,name={Fig.}, labelsep=period}
	\caption{Schematic diagram of RmMIMO-based multi-user downlink transmission in a single cell.}
	\label{Fig.scenario} 
	\vspace*{-1mm}
\end{figure}
We consider the downlink wideband transmission from an RmMIMO BS to $U$ single-antenna user equipments (UEs) in a time division duplexing (TDD) system, as shown in Fig. \ref{Fig.scenario}, while orthogonal frequency division multiplexing (OFDM) is adopted to combat frequency-selective fading. We assume that due to hardware constraints, each UE is equipped with an omnidirectional non-reconfigurable antenna, whereas the BS is equipped with a fully-digital uniform planar array (UPA)\footnote{Although this paper focuses on a fully-digital array-based RmMIMO architecture, the proposed precoding and channel estimation schemes can be easily extended to a hybrid antenna array with suitable algorithm modifications. We provide a comparison between the traditional fully-digital/hybrid array architectures and their reconfigurable conunterparts in Appendix \ref{apd1}.} comprising $M=M_yM_z$ antenna elements with $M_y$ and $M_z$ elements in the $y$- and $z$-direction, respectively. We assume that each BS antenna element avails of pattern reconfigurability. 

For the $u$-th $\left(1\le u \le U\right)$ UE, the received signal on the $g$-th subcarrier is given by
\begin{equation}\label{equ.ts}
	y_{u,g} =\bm{h}_{u,g}^{H}\bm{W}_g\bm{s}_g+n_{u,g}, 1\le g\le G,
\end{equation}
where $G$ is the total number of subcarriers, $\bm{h}_{u,g}^{H} = \left[h_{u,1,g}, h_{u,2,g}, \ldots, h_{u,M,g}\right] \in \mathbb{C}^{1\times M}$ is the downlink channel between the BS and the $u$-th UE; $\bm{W}_g \in \mathbb{C}^{M\times U}$ is the digital precoder; $\bm{s}_{g} \in \mathbb{C}^{U}$ denotes the transmitted signal on the $g$-th subcarrier such that $\mathbb{E}\left\{\bm{s}_{g}\bm{s}_{g}^{H}\right\} = \bm{I}_{U}$; and $n_{u,g} \sim \mathcal{N}_{c}\left(0, \sigma_{n}^2\right)$ is the complex additive white Gaussian noise (AWGN) at the UE.

\begin{table}[!h]
	\centering
	\captionsetup{name={TABLE}, font = {footnotesize}, justification = raggedright,labelsep=period}
	\caption{List of key parameters}    
	\resizebox{\columnwidth}{!}{
		\renewcommand\arraystretch{1.5}{
			{
				\begin{tabular}{c|l}
					\hline
					\textbf{Channel Parameters} & \multicolumn{1}{c}{\textbf{Definition}} \\ \hline
					$\tilde{x}_{i,u}$ & Channel gain of the $i$-th path \\ \hline
					$\left(\vartheta_{i,u},\varphi_{i,u}\right)$ & Zenith/azimuth angle of the $i$-th arrival path \\ \hline
					$\left(\theta_{i,u},\phi_{i,u}\right)$ & Zenith/azimuth angle of the $i$-th departure path \\ \hline
					${f}_{\rm{rx},u}\left(\vartheta,\varphi\right)$ & Radiation pattern of the $u$-th UE's antenna \\ \hline
					${f}_{\rm{tx},m}\left(\theta,\phi\right)$ & Radiation pattern of the $m$-th transmit antenna \\ \hline
					$\bm{k}_{\rm{tx},i,u}/\bm{k}_{\rm{rx},i,u}$ & Spherical unit vector of the $i$-th departure/$i$-th arrival path \\ \hline
					$\bm{p}_{m}/\bm{q}_{u}$ & Location vector of the $m$-th/$u$-th transmit/UE antenna \\ \hline
					$\lambda$ & Carrier wavelength \\ \hline
					$\tau_{i,u}$ & Channel delay of the $i$-th path \\ \hline
					$f_g$ & Frequency of the $g$-th subcarrier \\ \hline
					$L_u$ & Total number of multipath components \\ \hline
					$B_w$ & System bandwidth \\ \hline
					$G$ & Total number of subcarriers \\ \hline
					$M$ & Number of antennas of the BS array\\ \hline \hline
					\textbf{System Parameters} & \multicolumn{1}{c}{\textbf{Definition}} \\ \hline
					$h_{u,m,g}$ & sCSI for the $u$-th UE on $g$-th subcarrier at $m$-th antenna \\ \hline
					$\bm{q}_{u,m, g}$ & eCSI for the $u$-th UE on $g$-th subcarrier at $m$-th antenna\\ \hline
					$\bm{W}_{g}$ & Digital precoder on the $g$-th subcarrier  \\ \hline
					$\bm{\alpha}_{m}$ & EM domain precoder at BS's $m$-th antenna \\ \hline
					$\sigma_n^2$ & Variance of noise \\ \hline
					$J$  & Number of the pilot overhead in the frequency domain \\ \hline
					$T$  & Number of the pilot symbols in the time domain \\ \hline
					$K$  & Number of SH basis functions \\ \hline
					$U$  & Number of UEs \\ \hline
				\end{tabular}
			}
		}
	}
\end{table} 

\subsection{Channel Model}
The downlink channel between the $m$-th antenna and the $u$-th UE on the $g$-th subcarrier can be represented as follows\footnote{We consider a system with a relatively small bandwidth, such that the wavelengths of all subcarriers are approximately identical. For systems employing extremely large arrays and wide bandwidth, the beam squint effect \cite{BLW_TSP} for reconfigurable MIMO system design, channel estimation and precoding are interesting topics for further investigation in future work.}\cite{38.901}
\begin{align}
	\begin{split}
		\label{ch_full}
		h_{u,m,g} &= \sum_{i=1}^{L_u}\tilde{x}_{i,u} {f}_{{\rm{rx}},u}\left(\vartheta_{i,u},\varphi_{i,u}\right) {f}_{{\rm{tx}},m}\left(\theta_{i,u},\phi_{i,u}\right)\\&\times e^{-j\frac{2\pi }{\lambda}(\bm{k}_{{\rm{tx}},i,u}^{T}\bm{p}_{m} + \bm{k}_{{\rm{rx}},i,u}^{T}\bm{q}_{u} )}\cdot
		e^{-j2\pi\tau_{i,u}f_{g}}.
	\end{split}
\end{align}
Specifically, ${f}_{{\rm{tx}},m}\left(\theta_{i,u},\phi_{i,u}\right)$ is the radiation pattern gain of the $m$-th transmit antenna in direction $\left(\theta_{i,u},\phi_{i,u}\right)$, and  $\bm{k}_{{\rm{tx}},i,u}= \begin{bmatrix}\sin\theta_{i,u}\cos\phi_{i,u},\sin\theta_{i,u}\sin\phi_{i,u},\cos\theta_{i,u}\end{bmatrix}^{T}$ is the spherical unit vector of the $i$-th departure path according to the local coordinate system of the transmitter, as shown in Fig. \ref{Fig.scenario}. The position vector of the $m$-th BS antenna is defined as $\bm{p}_m = \left[0, \frac{2m_y+1-M_y}{2} d, \frac{M_z-2m_z-1}{2} d\right]$, where  $d$ is the antenna spacing, and the antenna index is given by $m = m_yM_z + m_z, 1\le m_y\le M_y, 1\le m_z\le M_z$, with $m_y$ and $m_z$ being the antenna index in $y$- and $z$-direction, respectively.
Furthermore, ${f}_{{\rm{rx}},u}\left(\vartheta_{i,u},\varphi_{i,u}\right)$ is the radiation pattern gain of the antenna of the $u$-th UE in direction $\left(\vartheta_{i,u},\varphi_{i,u}\right)$, $\bm{k}_{{\rm{rx}},i,u}$ is the spherical unit vector for the $i$-th arrival path at the receiver, and $\bm{q}_u$ is the position vector of the $u$-th UE's antenna. Furthermore, $f_g = \frac{g}{G}B_w, 1\le g \le G,$ is the frequency of the $g$-th subcarrier, where $B_w$ is the system bandwidth.\footnote{When the operating system bandwidth is not too large compared with the carrier frequency, the radiation pattern functions ${f}_{{\rm{rx}},u}\left(\vartheta,\varphi\right)$ and ${f}_{{\rm{tx}},m}\left(\theta,\phi\right)$ are approximately identical across all subcarriers \cite{RPA_BF, MRA_CE, AntennaTheory}.} 
Besides, $L_u$, $\tilde{x}_{i,u}$, $\tau_{i,u}$, and $\lambda$ are the number of multipath components, channel gain, channel delay, and carrier wavelength, respectively. 

By incorporating the delay term into the channel gain, i.e., $x_{i,u,g} = \tilde{x}_{i,u}e^{-j2\pi\tau_{i,u}f_g}$, we can represent the channel in a more compact form, that is 
\begin{equation}\label{hnm}
	h_{u,m,g} = \bm{f}_{{\rm{rx}},u}^{T}\bm{A}_{u}\bm{\mathit\Sigma}_{u,g}\bm{B}_{u,m}\bm{f}_{{\rm{tx}},u,m},
\end{equation}
where $\bm{f}_{{\rm{rx}},u} = [f_{{\rm{rx}},u}(\vartheta_{1,u},\varphi_{1,u}
),\ldots,f_{{\rm{rx}},u}(\vartheta_{L_u,u},\varphi_{L_u,u}
)]^{T}\in \mathbb{R}^{L_u}$, $\bm{f}_{{\rm{tx}},u,m} = [f_{{\rm{tx}},m}(\theta_{1,u},\phi_{1,u}
),\ldots,f_{{\rm{tx}},m}(\theta_{L_u,u},\phi_{L_u,u}
)]^{T}\in \mathbb{R}^{L_u}$. Moreover, we have the diagonal matrices $\bm{B}_{u,m} = \text{diag}\left(\bm{b}_{u,m}\right) \in \mathbb{C}^{L_u\times L_u}$ and $\bm{A}_{u} = \text{diag}\left(\bm{a}_{u}\right)\in \mathbb{C}^{L_u\times L_u}$ with $\bm{b}_{u,m} = \left[e^{-j\frac{2\pi}{\lambda}\bm{k}_{{\rm{tx}},1,u}^{T}\bm{p}_m},\ldots,e^{-j\frac{2\pi}{\lambda}\bm{k}_{{\rm{tx}},L_u,u}^{T}\bm{p}_{m}}\right]\in \mathbb{C}^{L_u}$ and $\bm{a}_u = \left[e^{-j\frac{2\pi}{\lambda}\bm{k}_{{\rm{rx}},1,u}^{T}\bm{q}_u},\ldots,e^{-j\frac{2\pi}{\lambda}\bm{k}_{{\rm{rx}},L_u,u}^{T}\bm{q}_{u}}\right]\in \mathbb{C}^{L_u}$, while $\bm{\mathit\Sigma}_{u,g} = \text{diag}\left(\left[{x}_{1,u,g}, x_{2,u,g}, \ldots, x_{L_u,u,g}\right]\right) \in \mathbb{C}^{L_u\times L_u}$ is the diagonal complex gain matrix. 
TmMIMO systems assume fixed radiation patterns, which limits the exploitation of the design DoF available in the EM domain. In this paper, we aim at designing transmit radiation patterns $\left\{{f}_{{\rm{tx}},m}\left(\theta,\phi\right)\right\}_{m=1}^{M}$ that match the signal propagation environment, such that the channel quality can be improved by the transmitter.

\subsection{Spherical Harmonious Orthogonal Decomposition Method}
Unfortunately, the transmit radiation pattern gain vectors $\left\{\bm{f}_{{\rm{tx}},u,m}\right\}_{m=1}^{M}$ interconnect the effects of the radiation pattern and the angles of departure (AoDs) in a nonlinear manner, which complicates the radiation pattern design. To overcome this challenge, we propose to use an orthogonal decomposition method that transforms the radiation pattern design problem to a projection coefficient vector optimization problem. 
We assume that a general radiation pattern can be linearly decomposed via $K$ orthogonal basis functions $\{\omega_{k}(\theta,\phi)\}_{k=1}^{K}$, i.e.,
\begin{equation}\label{equ.shod}
	f(\theta,\phi) = \sum_{k=1}^{K}\alpha_{k}\omega_{k}(\theta,\phi),
\end{equation}
where the basis functions satisfy the normalized orthogonal relationship, i.e., $	\iint \omega_{k}(\theta,\phi) \omega_{k^{\prime}}(\theta,\phi)\sin\theta d\theta d\phi = \left\{
\begin{aligned}
	0 \quad k\neq k^{\prime},\\
	1 \quad k = k^{\prime}.\\
\end{aligned}\right.$, and $\left\{\alpha_{k}\right\}_{k=1}^{K}$ are the weight coefficients. 

\begin{figure}[!t]
	\begin{center}
		\includegraphics[width = 1\columnwidth]{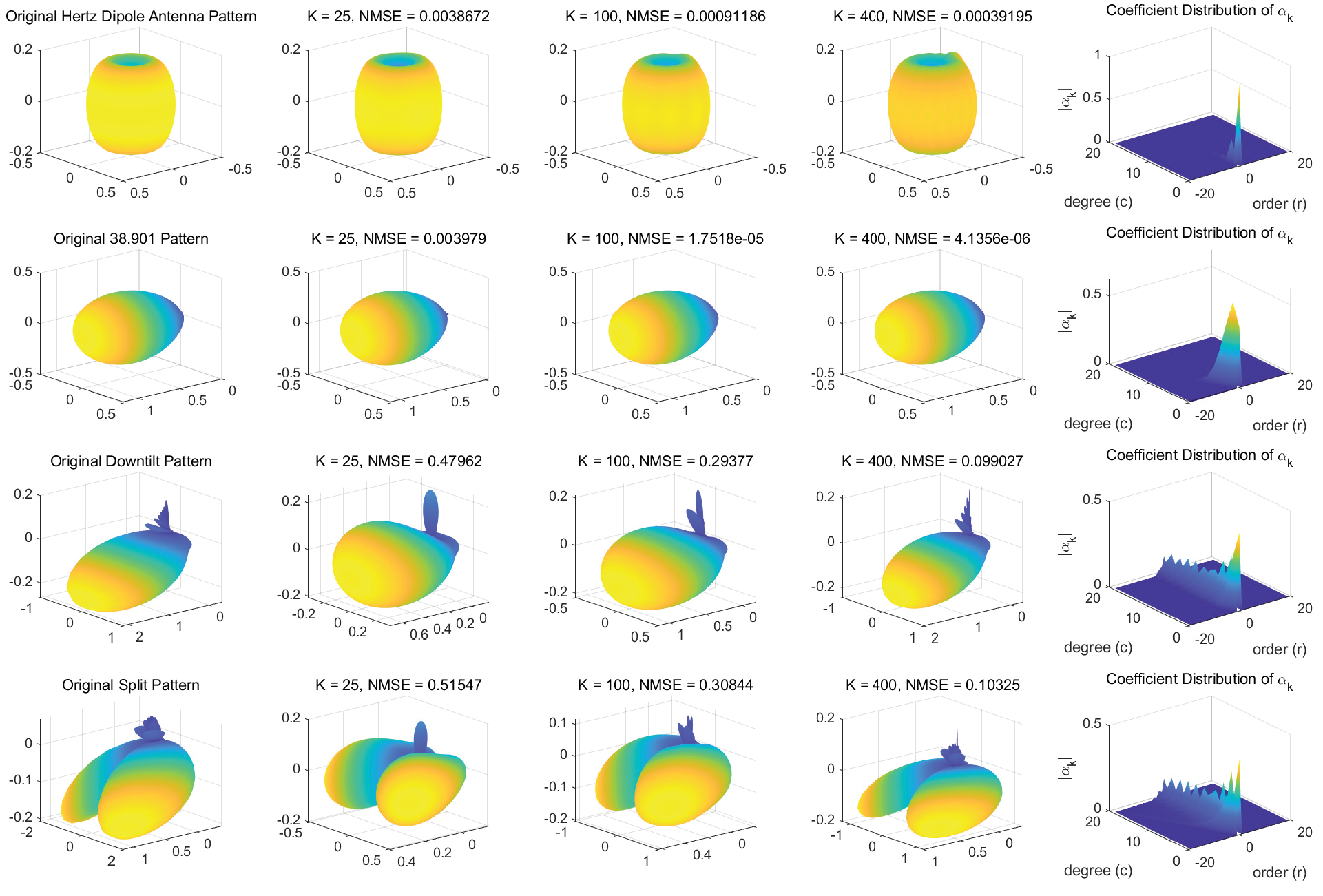}
	\end{center}
	\captionsetup{font={footnotesize}, singlelinecheck = off,name={Fig.}, labelsep=period}
	\caption{SHOD of four typical radiation patterns, where the original shape of the Hertz dipole antenna pattern, the 3GPP 38.901 pattern, a downtilt pattern, and a split pattern are compared with their respective reconstructed versions. Furthermore, the respective NMSEs and weight coefficients are shown. }
	\label{Fig.SHOD} 
	\vspace*{-5mm}
\end{figure}
Since we aim at optimizing the radiation pattern functions $\left\{{f}_{{\rm{tx}},m}\left(\theta,\phi\right)\right\}_{m=1}^{M}$, we wish to find a set of complete basis functions that can cover the whole two-dimensional $\left(\theta,\phi \right)$ space. Motivated by the application of spherical harmonious (SH) functions for planar wave expansion \cite{SH0}, array manifold decomposition \cite{SH}, and radiation pattern decomposition \cite{SH1}, we use SH functions to construct these bases, as they naturally satisfy the normalized orthogonal relationship in spherical coordinates. We refer to this approach as spherical harmonious orthogonal decomposition (SHOD) in this paper.\footnote{We note that other decomposition methods, such as two-dimensional (2D) discrete Fourier transform (DFT) bases, can also be considered. However, for the problem at hand, a 2D Fourier representation generally requires more coefficients than a spherical harmonic representation to achieve the same level of modeling accuracy \cite{SH}. Finding the optimal decomposition method remains an open question, and the proposed SHOD method can serve as a baseline for other radiation pattern decomposition schemes.
} Specifically, the SH functions parameterized by degree number $c$ and order number $r$ are denoted by $Y_{c}^r\left(\theta,\phi\right)$ \cite{SH}. Then, by defining the integer index $k = c^2 + c + r + 1, c \in [0,+\infty], r\in [-c,c]$, the $k$-th basis function is given by $\omega_{k}(\theta,\phi) = Y_{c}^r(\theta, \phi), k \in[0,+\infty]$,\footnote{While this paper focuses real-valued radiation pattern functions, by replacing the real spherical harmonics with complex ones \cite{SH}, the proposed method can be extended to complex-valued patterns. For fairness, since the baseline schemes from the literature we compare with (e.g. \cite{38.901}) are based on real-valued patterns, this paper also considers real-valued patterns.
} where
\begin{equation}\label{eq.ycr}
	Y_{c}^r(\theta, \phi) =\left\{\begin{matrix}(-1)^{r}\sqrt{2}K_c^r\cos(r\phi)P_c^r\left(\cos\theta\right)&0<r\le c,\\(-1)^{r}\sqrt{2}K_c^r\sin(-r\phi)P_c^{-r}\left(\cos\theta\right)&-c\le r<0,\\K_c^0P_c^r\left(\cos\theta\right)&r=0.\end{matrix}\right.
\end{equation}
Here, $P_c^r\left(\cos\theta\right)$ represents the Legendre function of degree $c$ and order $r$,  $K_c^r=\sqrt{\frac{2c+1}{4\pi}\cdot\frac{(c-|r|)!}{(c+|r|)!}}$ is a normalization factor, and $(c\pm|r|)!$ represents the factorial of $c\pm|r|$. 

To illustrate the potential of using SH functions as bases, we decompose four typical radiation patterns with different values of $K$: the Hertz dipole antenna radiation pattern, the 3GPP 38.901 radiation pattern \cite{38.901}, as well as a downtilt and a split radiation pattern generated by a reconfigurable parasitic antenna \cite{KeMag}.
We measured the reconstruction error using the normalized mean square error (NMSE) and show the distribution of the weight coefficients, $\{{\alpha}_k\}_{k=1}^{K}$, in Fig. $\ref{Fig.SHOD}$.
As can be seen, the reconstructed error becomes smaller as the number of basis functions $K$ increases. Besides, by considering the original radiation pattern shapes and the energy distribution of the weight coefficients, we can observe that most energy of the considered patterns concentrates in the low-frequency part, i.e., the positions with small $c$ and $r$. This is quite similar to the Fourier spectra of one-dimensional functions, where the low-frequency components contain the general shape information, while the high-frequency components capture the elaborate details. Different from the simple Hertz dipole antenna pattern and the 3GPP 38.901 pattern, the downtilt pattern and split pattern exhibit more complex high-frequency structure on their backside, which leads to the long tail of degree number $c$. Since most of the UEs are distributed at the broadside of the radiation pattern, the backside of the radiation pattern seldom affects the channel in practice. Therefore, with a sufficiently large value of $K$, we can accurately approximate the original radiation pattern.

Based on the above observations, we use the truncated SHOD to approximate the original radiation pattern gain vectors at the transmitter, i.e., 
\begin{equation}
	\bm{f}_{{\rm{tx}},u, m} = \bm{\mathit\Omega}_u\bm{\alpha}_{m}, \forall u, m,
\end{equation}
where the $\left(i,k\right)$-th element of $\bm{\mathit\Omega}_u \in \mathbb{R}^{L_u \times K}$ is given by $\left[\bm{\mathit\Omega}_u\right]_{i,k} = \omega_{k}\left(\theta_{i,u},\phi_{i,u}\right)$
and $\bm{\alpha}_{m} = [\alpha_{1,m}, \alpha_{2,m}, \ldots, \alpha_{K,m}]^{T}\in \mathbb{R}^{K}$ denotes the weight coefficient vector for the basis functions at the $m$-th antenna. Then, the channel coefficient in (\ref{hnm}) can be rewritten as follows
\begin{align}
	\begin{split}
		h_{u,m,g} = \bm{f}_{{\rm{rx}},u}^{T}\bm{A}_{u}\bm{\mathit{\Sigma}}_{u,g}\bm{B}_{u,m}\bm{\mathit\Omega}_u\bm{\alpha}_{m}  \overset{(a)}{=} \bm{q}_{u,m,g}^{H}\bm{\alpha}_{m},
	\end{split}	
\end{align}	
where in $(a)$ we introduce the definition $\bm{q}_{u,m,g}^{H} \triangleq \bm{f}_{{\rm{rx}},u}^{T}\bm{A}_{u}\bm{\mathit{\Sigma}}_{u,g}\bm{B}_{u,m}\bm{\mathit\Omega}_u\in \mathbb{C}^{1\times K}$, and $\bm{\alpha}_{m}$ is referred to as the EM domain precoder for the $m$-th antenna in the following. 
In this way, the original channel is decoupled into the product of a vector representing the environment $\bm{q}_{u,m,g}$, and a vector representing the radiation pattern $\bm{\alpha}_m$ at the transmitter. To differentiate between the newly introduced variable and the original channel, we refer to $\bm{q}_{u,m,g}$ as the eCSI and to $h_{n,m,g}$ as the sCSI. Since the eCSI captures the properties of the channel in the EM domain, exploiting it has the potential to improve the transmitter precoding design. In this study, we endeavor to design $\bm{\alpha}_m$ for maximum system throughput. Once an optimized $\bm{\alpha}_{m}$ is obtained, the radiation pattern of the $m$-th antenna can be reconstructed using (\ref{equ.shod}). Furthermore, the estimation of the newly introduced eCSI $\bm{q}_{u,m,g}$ will be discussed in Section \ref{S4}. Different from TmMIMO purely relying on digital domain precoding based on the sCSI, optimizing $\bm{\alpha}_m$ based on the eCSI can lead to an improved system throughput as the DoF in the EM domain are exploited.

\subsection{Problem Formulation}
Based on the proposed SHOD, the transmission model in (\ref{equ.ts}) can be rewritten as follows
\begin{equation}
	y_{u,g} = \bm{q}_{u,g}^{H}{\bm{\mathit\Lambda}}\bm{W}_{g}\bm{s}_{g}+n_{u,g}, 
\end{equation}
where $\bm{q}_{u,g} = \left[\bm{q}_{u,1,g}^{H}, \bm{q}_{u,2,g}^{H}, \ldots, \bm{q}_{u,M,g}^{H}\right]^{H} \in \mathbb{C}^{MK}$ is the aggregated eCSI vector for all transmit antennas, while $\bm{\mathit\Lambda} = \text{Blkdiag}\{\bm{\alpha}_{1},\bm{\alpha}_{2},\ldots, \bm{\alpha}_{M}\} \in \mathbb{C}^{MK\times M}$ is a block diagonal matrix with its diagonal blocks given by $\left\{\bm{\alpha}_{m}\right\}_{m=1}^{M}$. 
Then, the overall SE of the system is given by 
\begin{equation}\label{eq.se}
	R\! = \sum\limits_{g=1}^{G}\sum\limits_{u=1}^{U}\! \log_{2}\! \left(\!\! 1\! +\! \frac{|\bm{q}_{u,g}^{ H}{\bm{\mathit\Lambda}}\bm{w}_{u,g}|^2}{\sum\limits_{u^{\prime}\neq u}^{U}\! |\bm{q}_{u,g}^{H}{\bm{\mathit\Lambda}}\bm{w}_{u^{\prime},g}|^2\! +\! \sigma_n^2}\! \right)\!\! , \!
\end{equation}
where $\bm{w}_{u,g}$ is the digital precoder for the $u$-th UE, and $\bm{W}_{g} = \left[\bm{w}_{1,g}, \bm{w}_{2,g}, \ldots,  \bm{w}_{U,g}\right] \in \mathbb{C}^{M\times U}$. 

{Since the SH functions are normalized, the energy constraint for a given radiation pattern shape $f_{{\rm {tx}},m}\left(\theta,\phi\right)$ is equivalent to the energy constraint of the EM domain precoder $\|\bm{\alpha}_{m}\|^{2} = 1$. This can be validated by substituting (\ref{equ.shod}) into the normalized energy constraint $\iint |f(\theta,\phi)|^2 \sin\theta d\theta d\phi = 1$.} Thus, the SE optimization problem can be written as follows
\begin{align}\label{eq_problem}
	\begin{split}
		&\mathop{\max}\limits_{\{\bm{\alpha}_m\}_{m=1}^{M},\left\{\bm{W}_{g}\right\}_{g=1}^{G}} R \\ &
		\qquad\text{s.t.} \quad\|\bm{\alpha}_{m}\|^{2} = 1, \forall m, \\& \qquad\qquad\sum_{g=1}^{G}\|\bm{W}_g\|_{\mathrm{F}}^{2}\le P_T.
	\end{split}
\end{align}
Here, $P_T$ is the power constraint of the digital precoder. The SE optimization is non-convex in the design variable, which makes the joint optimization over the digital domain precoder $\left\{\bm{W}_{g}\right\}_{g=1}^{G}$ and EM domain precoder $\left\{\bm{\alpha}_{m}\right\}_{m=1}^{M}$ challenging. Inspired by the methods used to tackle traditional hybrid precoding problems \cite{MO, MO_weighted, MO_partially}, alternating optimization algorithms can be used due to their iterative nature and simplicity. Henceforth, we adopt an alternating optimization framework to solve problem (\ref{eq_problem}) in the following section, where the EM domain precoder is efficiently optimized through manifold optimization based on the eCSI, while the digital domain precoder is obtained through zero-forcing (ZF) based on the sCSI. Since this paper focuses on theoretically exploring the design of radiation patterns to enhance the system SE, the detailed hardware implementation of the EM domain precoder is beyond the scope of this study.

\section{Alternating Design of EM domain Precoder and Digital Domain Precoder}\label{S3}

\subsection{Single-Mode (SM) Design}
We first consider the SM optimization problem, i.e., all the BS transmit antennas employ the same radiation pattern. In this case, we have $\bm{\alpha}_{1} = \bm{\alpha}_{2} = \ldots = \bm{\alpha}_{M} = \bm{\alpha}$. Then, due to the block diagonal structure of $\bm{\mathit{\Lambda}}$, the overall SE in (\ref{eq.se}) can be rewritten as follows
\begin{equation}\label{cost}
	R\! = \sum\limits_{g=1}^{G}\sum\limits_{u=1}^{U}\! \log_{2}\! \left(\!\! 1\! +\! \frac{|\bm{\alpha}^{ H}\bm{Q}_{u,g}\bm{w}_{u,g}|^2}{\sum\limits_{u^{\prime}\neq u}^{U}\! |\bm{\alpha}^{H}\bm{Q}_{u,g}\bm{w}_{u^{\prime},g}|^2\! +\! \sigma_n^2}\! \right)\!\! , \!
\end{equation}  
where $\bm{Q}_{u,g} \triangleq \left[\bm{q}_{u,1,g}^{*}, \bm{q}_{u,2,g}^{*}, \ldots, \bm{q}_{u,M,g}^{*}\right] \in \mathbb{C}^{K\times M}$. By capitalizing on a method for solving traditional hybrid precoding problems \cite{MO}, the EM domain precoder $\bm{\alpha}$ and digital domain precoder $\left\{\bm{W}_g\right\}_{g=1}^{G}$ can be optimized alternatingly, where the non-convex constraint on $\bm{\alpha}$ can be effectively solved in the Riemannian manifold. 

We define the sphere manifold of the EM domain precoder as follows 
\begin{equation}
	\mathcal{S} = \{\bm{\alpha}\in \mathbb{R}^{K}: \|\bm{\alpha}\|_{2} = 1\}. 
\end{equation}
Then, the tangent space consisting of all the tangent vectors at point $\bm{\alpha}$ of $\mathcal{S}$ is denoted as $\mathcal{T}_{\bm{\alpha}}\mathcal{S} = \{\bm{z}\in \mathbb{R}^{K}: \bm{z}^{T}\bm{\alpha} = 0 \}.$
For all tangent vectors in the tangent space, the one representing the greatest decrease of the function is related to the Riemannian gradient. In the $\iota$-th iteration of the gradient descent, the Riemannian gradient can be obtained via the orthogonal projection of the Euclidean gradient $\nabla \mathfrak{f}(\bm{\alpha}_\iota)$ onto the tangent space $\mathcal{T}_{{\bm{\alpha}}_\iota}\mathcal{S}$, i.e.,
\begin{equation}\label{Rgrad}
	\text{grad}\mathfrak{f}(\bm{\alpha}_\iota) = \text{Proj}_{\bm{\alpha}_\iota}\nabla \mathfrak{f}(\bm{\alpha}_\iota) = \left(\bm{I}_{K}-\bm{\alpha}_\iota\bm{\alpha}_\iota^{T}\right)\nabla \mathfrak{f}(\bm{\alpha}_\iota).
\end{equation}
Here, $\text{Proj}_{\bm{\alpha}} = \left(\bm{I}_{K}-\bm{\alpha}\bm{\alpha}^{T}\right)$ is the projection operator for the sphere manifold, $\nabla = \frac{\partial }{\partial {\bm{\alpha}}}$ is the Euclidean gradient operator, $\mathfrak{f}\left({\bm\alpha}\right)$ is the objective function, and $\text{grad}\mathfrak{f}(\bm{\alpha})$ denotes the Riemannian gradient of the objective function. After determining the gradient direction in the tangent space, the updated point may no longer be on the Riemannian manifold. Therefore, the retraction operation is used to map the updated point back to the manifold, which can be represented as $\text{Retr}_{\bm{x}}: \mathcal{T}_{\bm{x}}\mathcal{S}\rightarrow \mathcal{S}$. The retraction operation in the $\iota$-th iteration is given by
\begin{equation}
	\bm{\alpha}_{\iota+1} = \text{Retr}_{\bm{\alpha}_\iota}\left(\gamma_\iota\bm{d}_{\iota}\right)  = \frac{\bm{\alpha}_\iota + \gamma_{\iota}\bm{d}_{\iota}}{\|\bm{\alpha}_\iota + \gamma_{\iota}\bm{d}_{\iota}\|_{2}},
\end{equation}
where $\gamma_{\iota}$ is the step size that can be determined through the Armijo backtracking line search \cite{MO} and $\bm{d}_\iota$ is the search direction. 
Based on the above definitions, we adopt the conjugate gradient algorithm to find the search direction, which is given by
\begin{equation}
	\bm{d}_{\iota} = -\text{grad}\mathfrak{f}\left(\bm{\alpha}_{\iota}\right) + \beta_{\iota}\text{Proj}_{\bm{\alpha}_{\iota}}(\bm{d}_{\iota-1}),
\end{equation}
where $\beta_{\iota}$ is the Polak-Ribiere parameter and $\text{Proj}_{\bm{\alpha}_{\iota}}(\bm{d}_{\iota-1})$ is the transportation operation which maps the last gradient descent direction into the current tangent space. 
For the single mode EM domain precoder optimization problem, the cost function is given by (\ref{cost}), i.e., $\mathfrak{f}(\bm{\alpha}_\iota) = -R_{\iota}$. Letting $\xi = \frac{1}{\sigma_{n}^2}$, the Euclidean gradient $\nabla \mathfrak{f}(\bm{\alpha}_\iota) = -\frac{\partial R_\iota}{\partial \bm{\alpha}_\iota} $ can be derived as follows	
\begin{align}\label{Egrad}
\nabla \mathfrak{f}({\bm{\alpha}}_\iota) = -\sum_{g=1}^{G}\sum_{u=1}^{U}\frac{\bm{\chi}^{\iota,\left(1\right)}_{u,g}-\bm{\chi}^{\iota, \left(2\right)}_{u,g}}{\ln 2},
\end{align} 
where $\bm{\chi}^{\iota,\left(1\right)}_{u,g}  \triangleq  \frac{2\xi \mathcal{R}e\left(\bm{Q}_{u,g}\bm{W}^{\iota}_{g}\left(\bm{W}^{\iota}_{g}\right)^{H}\bm{Q}_{u,g}^{H}\right)\bm{\alpha}_\iota}{1+\xi\|\bm{\alpha}_\iota^{H}\bm{Q}_{u,g}\bm{W}^{\iota}_{g}\|_2^{2}}$ and $\bm{\chi}^{\iota,\left(2\right)}_{u,g}  \triangleq  \frac{2\xi \mathcal{R}e\left(\bm{Q}_{u,g}\bm{W}^{\iota}_{\bar{u},g}\left(\bm{W}^{\iota}_{\bar{u},g}\right)^{H}\bm{Q}_{u,g}^{H}\right)\bm{\alpha}_\iota}{1+\xi\|\bm{\alpha}_\iota^{H}\bm{Q}_{u,g}\bm{W}^{\iota}_{\bar{u},g}\|_2^{2}}$. Specifically, $\bm{W}^{\iota}_{\bar{u},g} = \left[\bm{w}^{\iota}_{1,g}, \ldots, \bm{w}^{\iota}_{u-1,g}, \bm{w}^{\iota}_{u+1,g}, \bm{w}^{\iota}_{U,g}\right] \in \mathbb{C}^{M\times (U-1)}$ is composed of the precoding vectors for all $U$ UEs except for $\bm{w}^{\iota}_{u,g}$.

As described above, given the fixed digital domain precoder $\bm{W}_{g}^{\iota}, \forall g$, the EM domain precoder $\bm{\alpha}_{\iota+1}$ can be obtained through the conjugate gradient algorithm on the manifold. Subsequently, the digital domain precoder can be determined based on the sCSI once the EM domain precoder $\bm{\alpha}_{\iota+1}$ is known. 
According to the proposed channel model, the sCSI can be reconstructed as 
\begin{equation}
	\bm{h}_{u,g}^{\iota+1} = \bm{Q}_{u,g}^{H}\bm{\alpha}_{\iota+1}, \forall u,g.
\end{equation}
Then, exploiting the sCSI, traditional digital precoding algorithms can be applied, such as the well-known weighted minimum mean square error algorithm \cite{WMMSE}. In this paper, to keep the computational complexity low, we adopt the ZF precoder. Specifically, for the $u$-th UE, we first determine the normalized channel vector as $\bm{v}_{u,g}^{\iota+1} = \bm{h}_{u,g}^{\iota+1}/\|\bm{h}_{u,g}^{\iota+1}\|$. Then, the weight matrix of the digital precoder is given by
\begin{equation}\label{eq.ezf1}
	\tilde{\bm{W}}^{\iota+1}_{g} =  \bm{V}^{\iota+1}_{g}\left(\left({\bm{V}^{\iota+1}_{g}}\right)^{H}{\bm{V}^{\iota+1}_{g}}\right)^{-1},
\end{equation}
where $\bm{V}^{\iota+1}_{g} = \left[\bm{v}^{\iota+1}_{1,g},\bm{v}^{\iota+1}_{2,g}, \ldots, \bm{v}^{\iota+1}_{U,g}\right]\in \mathbb{C}^{M\times U}$ is the aggregated normalized sCSI vector matrix for all UEs. Finally, the digital precoder is normalized as 
\begin{equation}\label{eq.ezf2}
	\bm{W}^{\iota+1}_g = \sqrt{\frac{P_T}{G}}\frac{\tilde{\bm{W}}^{\iota+1}_g}{\|\tilde{\bm{W}}^{\iota+1}_g\|_{\text{F}}}.
\end{equation}

Based on the above procedure, we acquire the EM domain and digital domain precoders, respectively, as summarized in Algorithm \ref{alg1}, where the termination threshold $\eta_{\mathrm{th}}$ and maximum number of iterations $\iota_{\mathrm{max}}$ can be empirically chosen. To obtain a good initial solution, $\bm{\alpha}_0$ is initialized as the SH projection coefficient of a conventional radiation pattern, such as the dipole pattern. Then, the initialization of the digital precoder $\bm{W}^{0}_{g}$ can be obtained by applying ZF based on the corresponding sCSI $\bm{h}^{0}_{u,g} = \bm{Q}_{u,g}^{H}\bm{\alpha}_{0}$. Moreover, the convergence of the proposed manifold optimization procedure is guaranteed according to existing proofs in the literature \cite{BookMO}.

\begin{algorithm}[!t]
	\small
	\KwIn{eCSI $\{\bm{Q}_{u,g}\}_{u=1,g=1}^{U,G}$.} 
	\KwOut{EM precoder $\bm{\alpha}^{\star}$, digital precoder $\left\{\bm{W}_{g}^{\star}\right\}_{g=1}^{G}$.}
	Initialize the EM domain precoder $\bm{\alpha}_0$ as the SH projection coefficient of a dipole pattern, and let $\bm{h}_{u,g}^{0} = \bm{Q}_{u,g}^{H}\bm{\alpha}_0, \forall g$;\\
	Initialize the digital domain precoder $\left\{\bm{W}_{g}^{0}\right\}_{g=1}^{G}$ using  (\ref{eq.ezf1}) and (\ref{eq.ezf2});\\
	$\bm{d}_0 = -\text{grad}\mathfrak{f}(\bm{\alpha}_0)$; set iteration number to $\iota=0$;\\ 
	\Repeat{$|\frac{\mathfrak{f}(\bm{\alpha}_\iota)-\mathfrak{f}(\bm{\alpha}_{\iota-1})}{\mathfrak{f}(\bm{\alpha}_\iota)}|\le \eta_{\mathrm{th}}$ $\mathrm{or}$ ${\iota} \ge \iota_{\mathrm{max}}$} { 
		Choose the Armijo backtracking step size $\gamma_\iota$ \cite{MO}; \\	
		Update the next point $\bm{\alpha}_{\iota+1} = \text{Retr}_{\bm{\alpha}_\iota}\left(\gamma_\iota\bm{d}_{\iota}\right) $; \\
		Determine the Riemannian gradient $\bm{g}_{\iota+1} = \text{grad}\mathfrak{f}(\bm{\alpha}_{\iota+1})$ using (\ref{Rgrad}) and (\ref{Egrad}); \\
		Choose the Polak-Ribiere parameter as $\beta_{\iota+1} = \frac{\bm{g}_{\iota+1}^{H}\left(\bm{g}_{\iota+1}-\text{Proj}_{\bm{\alpha}_{\iota+1}}(\bm{g}_{\iota})\right)}{\|\text{Proj}_{\bm{\alpha}_{\iota+1}}(\bm{g}_{\iota})\|_2^{2}}$ \cite{MO}; \\
		Compute the conjugate direction $\bm{d}_{\iota+1} = -\bm{g}_{\iota+1} + \beta_{\iota+1}\text{Proj}_{\bm{\alpha}_{\iota+1}}(\bm{d}_{\iota})$; \\
		Construct the sCSI $\bm{h}_{u,g}^{\iota+1} = \bm{Q}_{u,g}^{H}\bm{\alpha}_{\iota+1}, \forall u$;\\
		Update the digital precoder $\bm{W}_{g}^{\iota+1}, \forall g$, using (\ref{eq.ezf1}) and (\ref{eq.ezf2});\\
		$\iota\leftarrow \iota+1$;\\
	}
	\Return $\bm{\alpha}^{\star} = \bm{\alpha}_{\iota}, \bm{W}_{g}^{\star} = \bm{W}_{g}^{\iota}, \forall g$.
	\caption{Alternating EM Domain and Digital Domain Precoder Design}
	\label{alg1} 
	\LinesNumbered	
\end{algorithm}
\subsection{Multi-Mode (MM) Design}
The above framework for SM optimization can be extended to MM optimization, where different antennas can adopt different radiation patterns. In this case, we need to return to optimization problem (\ref{eq.se}), where the EM domain precoder $\bm{\mathit{\Lambda}} = \text{Blkdiag}\{\bm{\alpha}_{1},\bm{\alpha}_{2},\ldots, \bm{\alpha}_{M}\}$ has to be optimized.  
The underlying problem is similar to the sub-connected hybrid precoding problem arising in traditional mMIMO systems, which can be addressed by modifying the manifold to the masked oblique manifold and re-deriving the corresponding Euclidean gradient \cite{MO_partially}.

We first consider the general oblique manifold. The oblique manifold is defined as $\mathcal{OB} = \{\bm{\mathit{\Lambda}}\in \mathbb{R}^{\tilde{K}\times M}: \left[\bm{\mathit{\Lambda}}^{T}\bm{\mathit{\Lambda}}\right]_{m,m} = 1, \forall m\}$, 
where $\tilde{K} = KM$. The tangent space of the oblique manifold defines a matrix $\bm{Z}\in \mathbb{R} ^{\tilde{K}\times M}$, whose columns are orthogonal to the corresponding columns of $\bm{\mathit{\Lambda}}$, i.e.,
\begin{equation}
	\mathcal{T}_{\bm{\mathit{\Lambda}}}\mathcal{OB} = \left\{\bm{Z}\in \mathbb{R}^{\tilde{K}\times M}: \left[\bm{\mathit{\Lambda}}^{T}\bm{Z}\right]_{m,m} = 0, \forall m\right\}.
\end{equation}
Similar to the sphere manifold case, the Riemannian gradient can be obtained by orthogonal projection of the Euclidean gradient $\nabla \mathfrak{f}(\bm{\mathit{\Lambda}}_\iota)$ onto the tangent space $\mathcal{T}_{\bm{\mathit{\Lambda}}_\iota}\mathcal{OB}$, i.e.,
\begin{align}\label{Rgrad2}
\begin{split}
\text{grad}\mathfrak{f}(\bm{\mathit{\Lambda}}_\iota) &= \text{Proj}_{\bm{\mathit{\Lambda}}_\iota}\nabla \mathfrak{f}(\bm{\mathit{\Lambda}}_\iota) \\&= \nabla \mathfrak{f}(\bm{\mathit{\Lambda}}_\iota) - \bm{\mathit{\Lambda}}_{\iota}\text{ddiag}\left(\bm{\mathit{\Lambda}}^{T}_\iota\nabla \mathfrak{f}(\bm{\mathit{\Lambda}}_\iota)\right),	
\end{split}
\end{align}
where $\text{ddiag}(\cdot)$ denotes the operator which sets all off-diagonal entries of a matrix to zero. The retraction operation is similarly defined as $\text{Retr}_{\bm{\mathit{\Lambda}}_\iota}: \mathcal{T}_{\bm{\mathit{\Lambda}}_\iota}\mathcal{OB}\rightarrow \mathcal{OB}$, and given by
\begin{equation}
	\bm{\mathit{\Lambda}}_{\iota+1} = \text{Retr}_{\bm{\mathit{\Lambda}}_\iota}\left(\gamma_\iota\bm{D}_{\iota}\right)  = \text{normalize}\left({\bm{\mathit{\Lambda}}_\iota + \gamma_{\iota}\bm{D}_{\iota}}\right),
\end{equation}
where $\text{normalize}(\cdot)$ scales each column of the input matrix to have norm 1. Adopting the conjugate gradient method, the gradient direction in the $\iota$-th iteration is given by
\begin{equation}
	\bm{D}_{\iota} = -\text{grad}\mathfrak{f}\left(\bm{\mathit{\Lambda}}_\iota\right) + \beta_{\iota}\text{Proj}_{\bm{\mathit{\Lambda}}_\iota}(\bm{D}_{\iota-1}),
\end{equation}
where $\text{Proj}_{\bm{\mathit{\Lambda}}_\iota}(\bm{D}_{\iota-1}) = \bm{D}_{\iota-1} - \bm{\mathit{\Lambda}}_{\iota}\text{ddiag}\left(\bm{\mathit{\Lambda}}^{T}_\iota\bm{D}_{\iota-1}\right) $. Next, we derive the Euclidean gradient $\nabla \mathfrak{f}(\bm{\mathit{\Lambda}}_\iota)$. By considering the block diagonal structure of $\bm{\mathit{\Lambda}}_\iota$, $\nabla \mathfrak{f}(\bm{\mathit{\Lambda}}_\iota)$ for the cost function $\mathfrak{f}(\bm{\mathit{\Lambda}}_\iota) =  -R_\iota $ is given by 
\begin{equation}
	\nabla \mathfrak{f}\left(\bm{\mathit{\Lambda}}_\iota\right)= \left( -\sum_{g=1}^{G}\sum_{u=1}^{U}\frac{ \bm{\Gamma}_{u,g}^{\iota,\left(1\right)} -\bm{\Gamma}_{u,g}^{\iota,\left(2\right)}}{\ln 2}
\right) \odot \bm{M}_{0},
\end{equation}
where $\bm{\Gamma}_{u,g}^{\iota,\left(1\right)} \triangleq \frac{2\xi \mathcal{R}e\left(\bm{q}_{u,g}\bm{q}_{u,g}^{H}\bm{\mathit{\Lambda}}_\iota\bm{W}_g^{\iota}\left(\bm{W}_g^{\iota}\right)^{H}\right)}{1+\xi\|\bm{q}_{u,g}^{H}\bm{\mathit{\Lambda}}_\iota\bm{W}_{g}^{\iota}\|_2^{2}}$,  $\bm{\Gamma}_{u,g}^{\iota,\left(2\right)} \triangleq \frac{2\xi \mathcal{R}e\left(\bm{q}_{u,g}\bm{q}_{u,g}^{H}\bm{\mathit{\Lambda}}_\iota\bm{W}_{\bar{u},g}^{\iota}\left(\bm{W}_{\bar{u},g}^{\iota}\right)^{H}\right)}{1+\xi\|\bm{q}_{u,g}^{H}\bm{\mathit{\Lambda}}_\iota\bm{W}_{\bar{u},g}^{\iota}\|_2^{2}}$, and $\bm{M}_0$ is a mask block diagonal matrix defined as $\bm{M}_{0} = \text{Blkdiag}\{\bm{m}_{1},\ldots, \bm{m}_{M}\}\in\mathbb{R}^{\tilde{K}\times M}$, and  $\bm{m}_{1} = \bm{m}_{2} = \ldots = \bm{m}_{M} = \bm{1}_{K}$.

Given the above definitions of the oblique manifold, tangent space, retraction operation, gradient direction, and Euclidean gradient, the MM EM precoder $\bm{\mathit{\Lambda}}$ and digital precoder $\bm{W}_g, \forall g$, can also be alternatingly optimized similar to the procedure in Algorithm \ref{alg1}. Compared with the SM case, where all the antennas employ the same radiation pattern, the MM case enables the joint design of multiple antennas' EM radiation patterns, thereby offering more DoF for enhanced SE performance.

\subsection{Complexity Analysis}
In this section, we analyze the computational complexity of the proposed schemes for the SM and MM cases, respectively.

\subsubsection{SM Complexity}
According to Algorithm \ref{alg1}, the complexity mainly stems from the following three parts. 

$\bullet$ Armijo backtracking line search in line 5: The step size is iteratively searched with exponential decay in the Armijo backtracking line search, and the main computation therein comes from evaluating the cost function $ f\left(\bm{\alpha}_{\iota}\right)$, leading to an approximate complexity of $\mathcal{O}\left(N_{\textit{ls}}\left(GMUK+GMU^2\right)\right)$. Here, $N_{\textit{ls}}$ denotes the number of search steps, which empirically does not exceed $10$-$20$.  

$\bullet$ Computation of the Riemannian gradient in line 7: The matrix multiplications in (\ref{Egrad}) contribute most to the overall complexity in this line. By neglecting the terms that are far less than the others, the complexity is given by $\mathcal{O}\left(2GU\left(M^2K+M^2U+K^2M\right)\right)$. 

$\bullet$ Complexity of digital domain precoder computation in lines 10-11: The computational complexity of ZF-based digital precoding is given by $\mathcal{O}\left(2GMU^2+\frac{2}{3}GU^3 + 2GMU+GMUK\right)$. Here, the matrix inversion and multiplication operations contribute most of the computations.

Furthermore, since typical values for the number of users, $U$, are much smaller than typical values for the number of BS antennas, $M$, and the number of basis functions, $K$, we simplify the complexity of the above three parts as $\mathcal{O}\left(N_{\textit{ls}}GMUK\right)$, $\mathcal{O}\left(2GU\left(M^2K+K^2M\right)\right)$, and $\mathcal{O}\left(GMUK\right)$, respectively. To sum up, retaining only the high-order terms, the overall complexity of the proposed SM precoding design is given by $\mathcal{O}\left(\left(2K+2M+N_{ls}\right)GMUKN_{iter}\right)$, where $N_{iter}$ is the number of iterations of Algorithm \ref{alg1}.

\subsubsection{MM Complexity}
The approximate computational complexity of the Armijo backtracking line search and the computation of the digital domain precoder are the same as in the SM case. For the computation of the Riemannian gradient in the MM case, the complexity is approximately $\mathcal{O}\left(2GU\left(2K^2M^2+M^2U+KM^3\right)\right)$, which can be further simplified as $\mathcal{O}\left(2GU\left(2K^2M^2+KM^3\right)\right)$. Therefore, the overall complexity of the proposed MM precoder design is $\mathcal{O}\left(\left(4KM+2M^2+N_{ls}\right)GMUKN_{iter}\right)$.

\begin{figure*}
	\hspace{-5mm}
	\subfigure[]{
		\begin{minipage}[t]{0.25\linewidth}
			\centering
			\includegraphics[width=\textwidth]{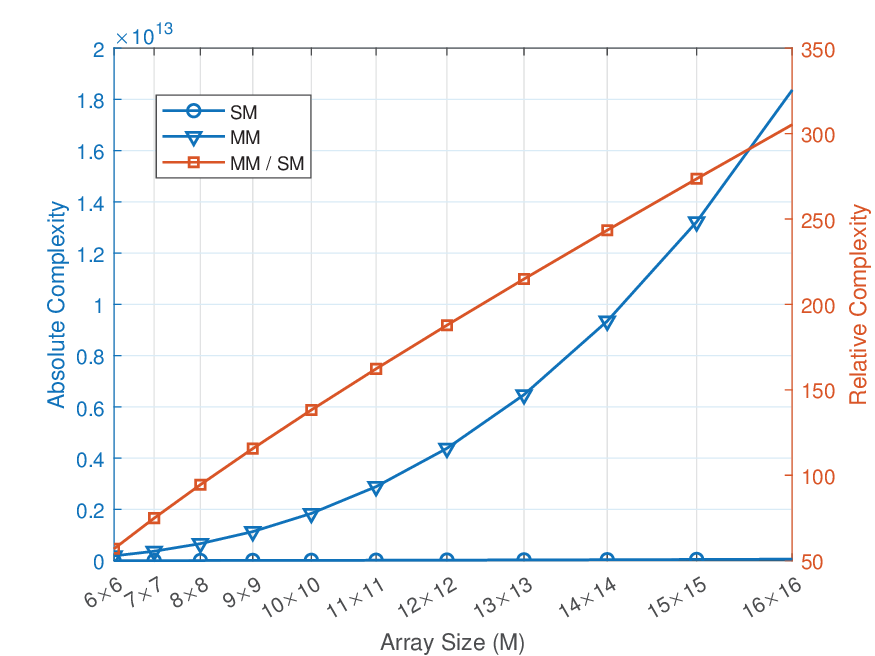}
		\end{minipage}
	}%
	\hspace{-4mm}
	\subfigure[]{
		\begin{minipage}[t]{0.25\linewidth}
			\centering
			\includegraphics[width=\textwidth]{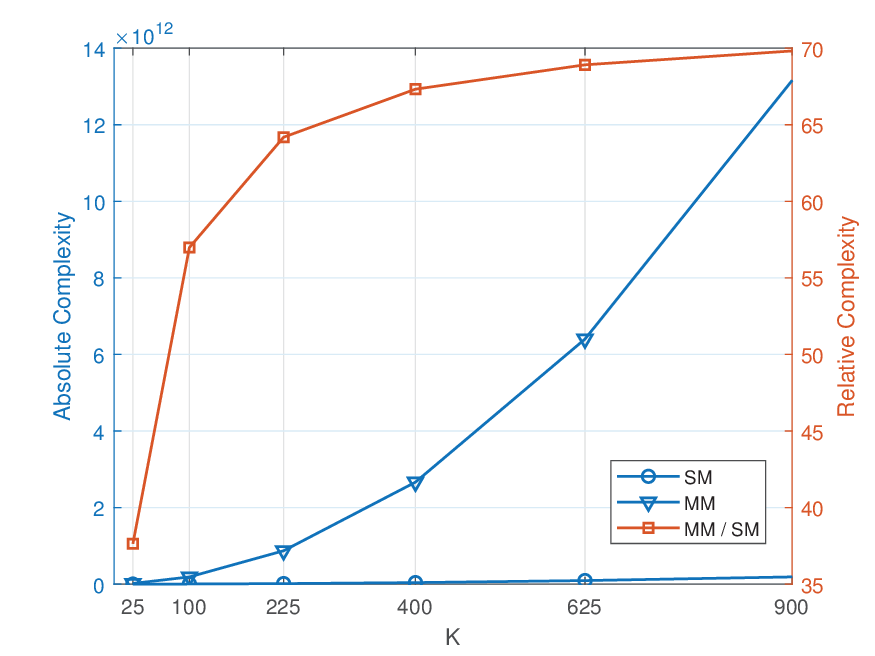}
		\end{minipage}
	}%
	\hspace{-4mm}
	\subfigure[]{
		\begin{minipage}[t]{0.25\linewidth}
			\centering
			\includegraphics[width=\textwidth]{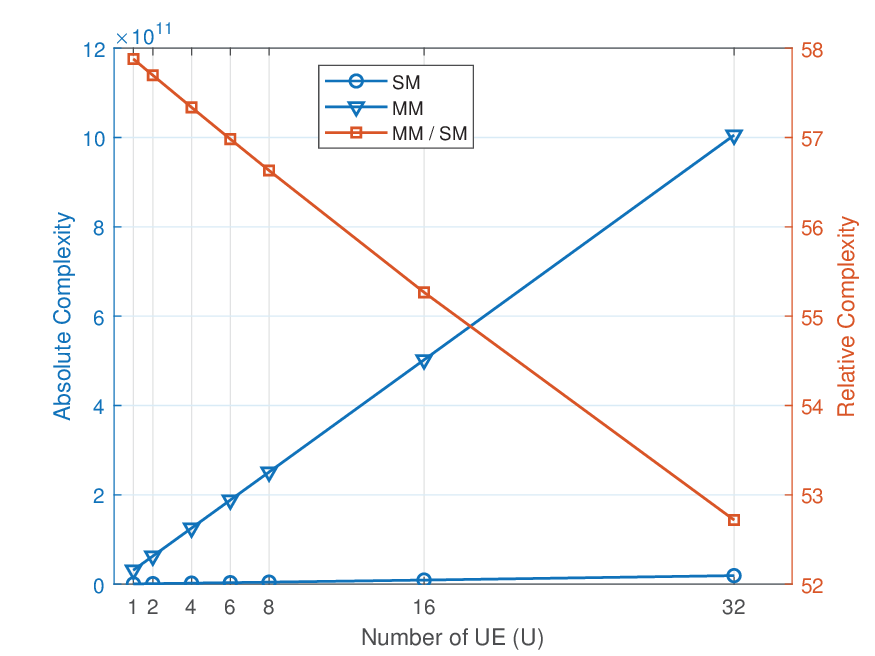}
		\end{minipage}
	}%
	\hspace{-4mm}
	\subfigure[]{
		\begin{minipage}[t]{0.25\linewidth}
			\centering
			\includegraphics[width=\textwidth]{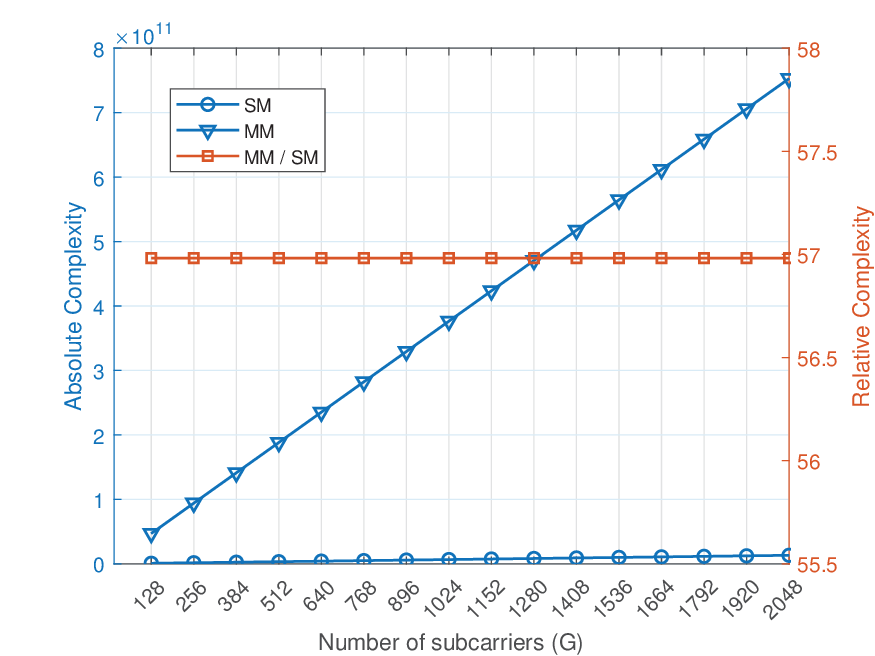}
		\end{minipage}
	}%
	\vspace*{-2mm}  
	\captionsetup{font={footnotesize}, singlelinecheck = off,name={Fig.}, labelsep=period}
	\caption{a) Complexity versus the array size $M$, b) complexity versus the number of SH functions $K$, c) complexity versus the number of UEs $U$, d) complexity versus the number of subcarriers $G$.}
	\label{fig_complexity_pre}
	\vspace*{-5mm}
\end{figure*}

\subsubsection{Complexity Comparison}
In Figs. \ref{fig_complexity_pre} (a)-(d), we plot the number of complex multiplications of the precoding algorithm for the SM and MM designs, as well as the ratio of the MM/SM computational complexity, as functions of the key algorithm parameters.
Figure \ref{fig_complexity_pre} (a) shows that the computational complexities of SM and MM precoding increase quadratically and cubically with the number of antennas, respectively. Thus, as the number of antennas increases, the computational complexity of MM becomes significantly higher than that of SM.
Figure \ref{fig_complexity_pre} (b) shows that the computational complexities of SM and MM grow approximately quadratically with the number of SH functions $K$. As $K$ increases, the computational complexity of MM becomes higher than that of SM, but the ratio of their complexities approaches a constant value. Moreover, Fig. \ref{fig_complexity_pre} (c) and Fig. \ref{fig_complexity_pre} (d) indicate that the computational complexities of SM and MM both increase approximately linearly with both the number of UEs and  the number of subcarriers.

These results suggest that the primary parameters affecting the algorithm complexity are the number of antennas $M$ and the number of the SH functions $K$. Besides, MM precoding always entails much higher complexity than SM precoding. Additionally, although digital precoding involves matrix inversions, its impact on the overall computational complexity is not substantial since the dimension of inverse matrices (the number of UEs) is comparatively small.

\section{Channel Estimation in the Electromagnetic Domain}\label{S4}
\begin{figure}[!t]
	\begin{center}
		\includegraphics[width = 1\columnwidth]{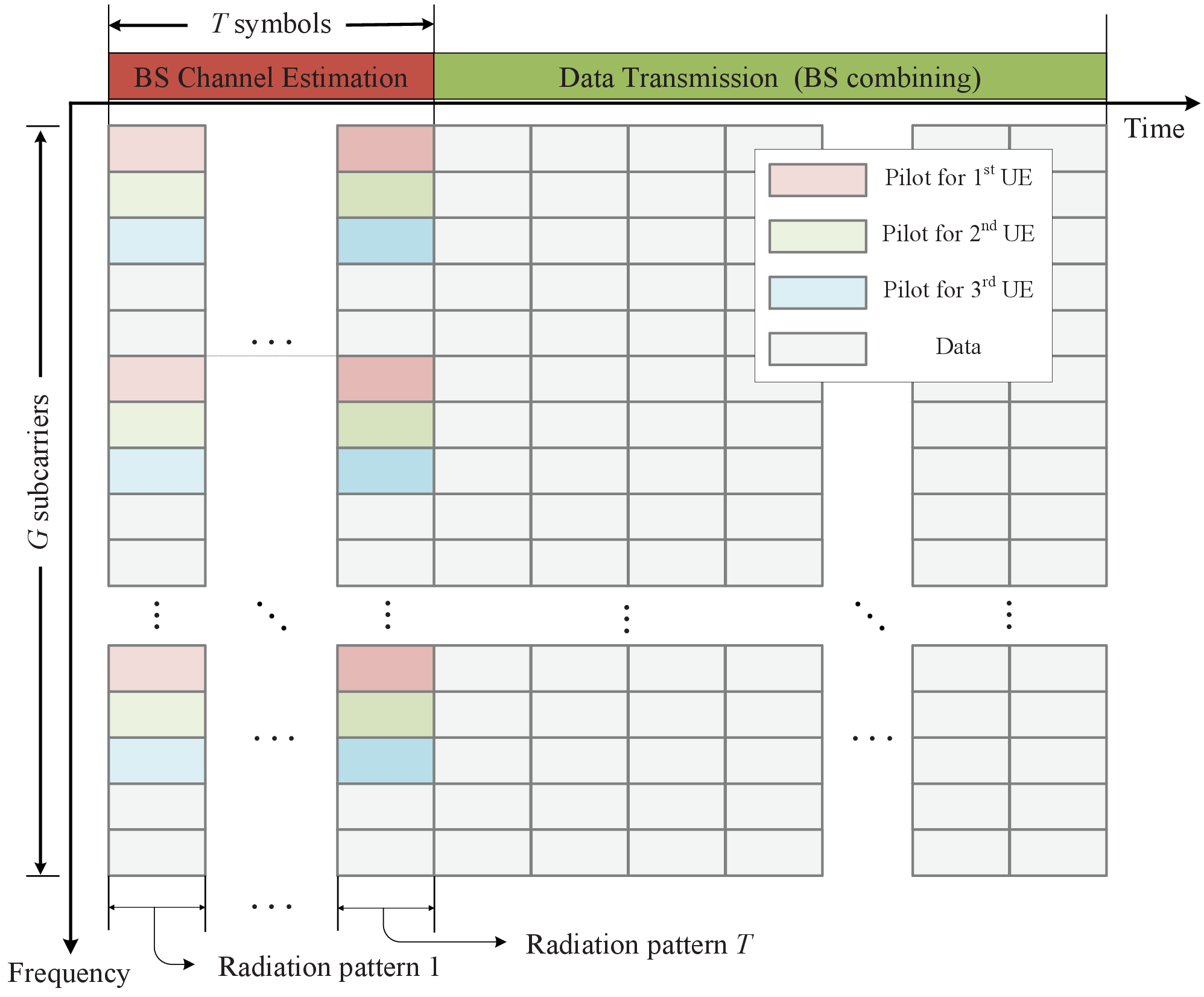}
	\end{center}
	\captionsetup{font={footnotesize}, singlelinecheck = off,name={Fig.}, labelsep=period}
	\caption{Transmission frame structure of the proposed uplink channel estimation and combing scheme.}
	\label{Fig.Frame} 
	\vspace*{-5mm}
\end{figure}

For the proposed single/multi-mode precoding design, knowledge of the eCSI $\bm{q}_{u,m,g},\forall u,m,g$, is required. In the context of TmMIMO systems, the acquisition of the sCSI ${h}_{u,m,g}$ has been well-researched. Notwithstanding, there is a dearth of previous research specifically focusing on acquiring $\bm{q}_{u,m,g}$. In this section, we first propose a subspace-based sCSI estimation method for reducing the frequency domain pilot overhead during uplink transmission. Then, we extend this method to the EM domain for multi-carrier eCSI reconstruction. In this way, the estimated uplink eCSI can be used for BS combiner design. {The transmission frame structure of the proposed uplink channel estimation and combining procedure is shown in Fig. \ref{Fig.Frame}. Since we consider a TDD system, the precoding scheme employed by the BS during downlink transmission is identical to the combiner design for uplink transmission.}

\subsection{sCSI Estimation}\label{Sec-sCSI}
We first address the wideband sCSI estimation for multi-user uplink transmission, where we aim to estimate the full-band sCSI from limited undersampled frequency observations. Specifically, we assign different subcarriers to different users to avoid interference among the users. Additionally, we leverage the sparsity property of the delay domain channel to reduce the pilot overhead in the frequency domain. 

\subsubsection{Problem Formulation}
We consider the uplink sCSI estimation between the BS and $U$ single-antenna UEs. Consistent with the configurations mentioned in Section \ref{S3.2}, the BS employs a fully digital UPA comprising $M$ antennas. Among all the $G$ subcarriers in the frequency domain, each UE selects a subset $\mathcal{G}_u = \left\{g_{u,1}, g_{u,2}, \ldots, g_{u,J}\right\} \subset \left\{1,2,\ldots, G\right\}$ for transmitting pilot signals. The selected subsets for any two distinct UEs are disjoint and of equal cardinality, specifically, $\mathcal{G}_{u} \cap \mathcal{G}_{u^{\prime}} = \varnothing, |\mathcal{G}_{u}| =  |\mathcal{G}_{u^{\prime}}| = J, \forall u\neq u^{\prime}$. In the time domain, $T$ continuous pilot symbols are employed for each UE. Here, we assume that the channel parameters remain constant during channel estimation and the subsequent data transmission.

Since the pilot signals of different users are orthogonal in the frequency domain, we consider the uplink sCSI estimation between a specific UE $u$ and the BS. To simplify notation, we omit subscript $u$ in the following. Then, the $t$-th symbol $\left(1\le t\le T\right)$ received by the $m$-th antenna $\left( 1\le m \le M \right)$ of the BS on the $g_{j}$-th subcarrier $\left(1\le j \le J \le G/U \right)$ can be expressed as
\begin{equation}
	\begin{aligned}\label{equ.ce_sCSI}
		{y}_{m,g_{j}}^{\left(t\right)} = \left({h}_{m,g_j}^{\left(t\right)}\right)^{*} s_{g_j}^{\left(t\right)} + {n}_{m,g_j}^{\left(t\right)} = \left(\bm{\alpha}_{m}^{\left(t\right)}\right)^{T}\bm{q}_{m,g_j}s_{g_j}^{\left(t\right)} + n_{m,g_j}^{\left(t\right)}.
	\end{aligned}
\end{equation}
Here, the uplink channel $\left(h_{m,g_j}^{\left(t\right)}\right)^{*}$ is time-dependent since the BS can alter the receive radiation pattern via $\bm{\alpha}_{m}^{\left(t\right)}$ over time to provide diverse observations. Note that $\bm{\alpha}_{m}^{\left(t\right)}$ is frequency-independent, since the channels at different subcarriers share the same radiation pattern within the system bandwidth. Nevertheless, the eCSI $\bm{q}_{m,g_l}$ is dependent on the antenna index and the subcarrier. For the sCSI recovery problem, we need to recover the full band sCSI $h_{m,g}^{\left(t\right)}, \forall m,g,t$ based on the undersampled observations $y_{m,g_j}^{\left(t\right)}, \forall m,g_j,t$, and the known pilots $s_{g_j}^{\left(t\right)}, \forall g_j, t$.

\subsubsection{1D-ESPRIT}
Since the same procedure can be consistently applied to the sCSI estimation at different time instances $t$, we concentrate on one specific pilot symbol and omit the subscript $t$ in the following. Thus, we have ${y}_{m,g_j} = {h}_{m,g_j}^{*} s_{g_j} + {n}_{m,g_j}$. 
Define $\bm{h}_{m}^{*} \triangleq \left[{h}_{m,1}^{*}, \ldots, {h}_{m,g}^{*}, \ldots, {h}_{m,G}^{*}\right]^{T} \in \mathbb{C}^{G}$ as the full-dimensional frequency domain sCSI to be estimated. According to the channel model in (\ref{ch_full}), the conjugate of the $g$-th element of $\bm{h}_{m}^{*}$ can be rewritten as 
\begin{align}\label{ch_dd}
	\begin{split}	
		h_{m,g} \triangleq\sum_{i=1}^{L_u}\bar{x}_{m,i}e^{-j2\pi\tau_{i}f_{g}}, 
	\end{split}
\end{align}
where $\bar{x}_{m,i}\!\! \triangleq \!\! \tilde{x}_{i} {f}_{\rm{rx}}\left(\vartheta_{i},\varphi_i\right) {f}_{{\rm{tx}},m}\left(\theta_{i},\phi_i\right)e^{-j\frac{2\pi }{\lambda}(\bm{k}_{{\rm{tx}},i}^{T}\bm{p}_{m} + \bm{k}_{{\rm{rx}},i}^{T}\bm{q})}$ is the equivalent frequency domain channel gain. 

Next, we introduce a subspace-based sCSI estimation method, which provides super-resolution estimates of the channel delay. Based on the considered transmission model and (\ref{ch_dd}), the observations can be rewritten as 
\begin{equation}\label{ce:sCSI3}
	\bm{y}_m = \bm{S}\bm{E}_d\bm{x}_{m} + \bm{n}_m,
\end{equation}
where $\bm{y}_m = \left[{y}_{m,g_1}, {y}_{m,g_2}, \ldots, {y}_{m,g_J}\right]^{T}\in\mathbb{C}^{J}$ is the undersampled frequency domain observation, and $\bm{n}_m = \left[{n}_{m,g_1}, {n}_{m,g_2}, \ldots, {n}_{m,g_J}\right]^{T}\in\mathbb{C}^{J}$ is the noise vector. Furthermore, $\bm{x}_m = \left[\bar{x}_{m,1}^{*}, \bar{x}_{m,2}^{*}, \ldots, \bar{x}_{m,L_u}^{*} \right]^{T} \in \mathbb{C}^{L_u}$ contains the equivalent frequency domain channel gains, $\bm{S} = \text{diag}\left(\left[s_{g_1}, s_{g_2}, \ldots, s_{g_J}\right]\right)\in \mathbb{C}^{J\times J}$, and $\bm{E}_d = \left[ \bm{e}_{d}^{1}, \bm{e}_{d}^{2}, \ldots, \bm{e}_{d}^{L_u} \right]\in \mathbb{C}^{J\times L_u}$ with $\bm{e}_d^{i} = \left[e^{j2\pi\tau_{i}f_{g_1}}, e^{j2\pi\tau_{i}f_{g_2}}, \ldots, e^{j2\pi\tau_{i}f_{g_J}} \right]^{T} \in \mathbb{C}^{J}, \forall i \in \left\{1,2, \ldots, L_u\right\}$.
By stacking $M$ observations obtained from different antennas and multiplying both sides of (\ref{ce:sCSI3}) by the inverse of matrix $\bm{S}$, we have 
\begin{equation} \label{ce:sCSI4}
\bm{Y}_d = \bm{E}_d\bm{X}_d + \bm{N}_d,
\end{equation}
where $\bm{Y}_d = \bm{S}^{-1}\left[\bm{y}_1,\bm{y}_2,\ldots,\bm{y}_M\right] \in \mathbb{C}^{J\times M}$, $\bm{X}_d = \left[\bm{x}_1,\bm{x}_2,\ldots,\bm{x}_M \right]\in \mathbb{C}^{L_u\times M}$, and $\bm{N}_d = \bm{S}^{-1}\left[\bm{n}_1, \ldots, \bm{n}_M\right] \in \mathbb{C}^{J\times M}$.

Therefore, by assigning pilot sequences with uniform spacing in the frequency domain, the phase of the elements in each column of $\bm{E}_d$ forms an arithmetic progression. To achieve higher delay resolution, the selected subcarriers should have a large frequency span. For a $G$ subcarrier system with $J$ pilot symbols in the frequency domain, the $j$-th pilot can be uniformly inserted at frequency $f_{g_j} = f_{g_1} + \lfloor\frac{G}{J}\rfloor (j-1)\Delta f, j\in \left\{1,\ldots,J\right\}$. Here, $\Delta f = \frac{B_w}{G}$ is the subcarrier spacing, and $f_{g_{1}}$ corresponds to the first subcarrier allocated to the considered user. Thus, the phase difference, $\kappa_{i} \triangleq 2\pi\tau_{i}\lfloor\frac{G}{J}\rfloor \Delta f, \forall i$, between two adjacent elements of the vector $\bm{e}_d^{i}$ can be effectively estimated using the 1D-ESPRIT algorithm, and the number of multipath components $L_u$ can be effectively estimated through an eigenvalue decomposition of $\bm{Y}_d\bm{Y}_d^{H}$\cite{Liao_Tcom}.

By applying the 1D-ESPRIT algorithm, we can acquire the estimates of $\left\{\hat{\kappa}_{i}\right\}_{i=1}^{L_u}$, from which the estimates of the multipath delays can be obtained as follows:
\begin{equation}\label{ce:tau}
	\hat{\tau_{i}} = \frac{\hat{\kappa_{i}}}{2\pi\lfloor\frac{G}{J}\rfloor \Delta f}, \forall i.
\end{equation} 
Based on the estimated multipath delays $\left\{\hat{\tau_{i}}\right\}_{i=1}^{L_u}$, we can acquire an estimate for $\hat{\bm{E}}_d$ by substituting $\left\{\hat{\tau}_{i}\right\}_{i=1}^{L_u}$ back into the definition of ${\bm{E}}_d$. Then, the multipath channel gain can be determined by applying LS estimation based on (\ref{ce:sCSI4}), i.e.,
\begin{equation}\label{ce:ls1}
	\hat{\bm{X}}_d = \left(\hat{\bm{E}}_d^{H}\hat{\bm{E}}_d\right)^{-1}\hat{\bm{E}}_d^{H}\bm{Y}_d.
\end{equation}
With the estimated channel gains $\left\{\hat{\bm{x}}_m\right\}_{m=1}^{M}$ and multipath delays $\left\{\hat{\tau}_{i}\right\}_{i=1}^{L_u}$, the sCSI on any subcarrier can be reconstructed by substituting the estimated parameters back into (\ref{ch_dd}), i.e., $\hat{h}_{m,g} = \sum_{i=1}^{L_u}\hat{x}_{m,i}e^{-j2\pi\hat{\tau}_{i}f_{g}}, \forall m,g$.
For practical implementation, we assume that the channel parameters remain constant during both channel estimation and subsequent data transmission. As a result, we reintroduce the time index $t$ into (\ref{ce:sCSI4}), the $T$ continuous time domain received pilot symbols $ \bm{Y}_d^{\left(t\right)}$, $ t \in \{1, 2, \dots, T\} $, can be stacked by columns. Subsequently, the 1D-ESPRIT algorithm can be applied to enhance the delay estimation robustness against noise.
Since the core idea of this sCSI estimation method is based on ESPRIT in the frequency domain, we hereafter refer to it as the 1D-ESPRIT algorithm.

\subsection{eCSI Estimation} \label{Sec-eCSI}
Having obtained the wideband sCSI in Section \ref{Sec-sCSI}, now we focus on estimating the eCSI. Since all users employ the same estimation procedure, we concentrate on solving the single-user eCSI acquisition problem also in this case.  

\subsubsection{Problem Formulation}
Recalling (\ref{equ.ce_sCSI}), the signal model is given by ${y}_{m,g_j}^{\left(t\right)} = \left({h}_{m,g_j}^{\left(t\right)}\right)^{*} s_{g_j}^{\left(t\right)} + {n}_{m,g_j}^{\left(t\right)} = \left(\bm{\alpha}^{\left(t\right)}\right)^{T}\bm{q}_{m,g_j}s_{g_j}^{\left(t\right)} + n_{m,g_j}^{\left(t\right)}.$
For the eCSI recovery problem, we need to recover the full-band eCSI $\bm{q}_{m,g}, \forall m,g,$ based on the observations $y_{m,g_j}^{\left(t\right)}, \forall m,g_j,t$, known pilots $s_{g_j}^{\left(t\right)}, \forall g_j, t$, known pattern projection coefficients $\bm{\alpha}_{m}^{\left(t\right)}, \forall t$, and estimated sCSI $\hat{h}_{m,g}^{\left(t\right)}, \forall m,g,t$. 
For simplicity, we assume that all antennas employ the same radiation pattern at time $t$ of the channel estimation stage. Therefore, $\bm{\alpha}^{\left(t\right)}$ is not dependent on the antenna index $m$.
One straightforward method to estimate the eCSI is to stack $T$ consecutive received pilot symbols. Then, the eCSI can be estimated through the simple least squares method when $T\ge K$. However, such a method can incur high pilot overhead when the dimension of $\bm{q}_{m,g_{j}}^{\left(t\right)}$ is large. On the other hand, there is no obvious sparsity in the eCSI. Therefore, the traditional compressed sensing methods are not directly applicable.

To overcome the above challenges, we need to exploit the structure information of the eCSI, i.e., $\bm{q}_{m,g}^{H} = \bm{f}_{\rm{rx}}^{T}\bm{A}\bm{\mathit\Sigma}_{g}\bm{B}_{m}\bm{\mathit\Omega}$. Here, the subscript $u$ has been omitted for notational simplicity. Specifically, we divide the eCSI estimation problem into two stages. In the first stage, the multipath AoAs at the BS side are estimated; based on the estimated AoAs, the matrices $\bm{B}_m$ and $\bm{\mathit\Omega}$ can be constructed. In the second stage, we estimate the equivalent channel gain and recover the entire eCSI. Here, we define $\bm{r}_g \triangleq \left( \bm{f}_{\rm{rx}}^{T}\bm{A}\bm{\mathit\Sigma}_g\right)^{H} \in\mathbb{C}^{L_u}$, which can be seen as an equivalent channel gain vector that incorporates the effects of the channel gain, delay, subcarrier, and AoD information at the UE side. 

\subsubsection{2D-ESPRIT} 
Based on (\ref{equ.ce_sCSI}), the transmission model can be rewritten as
\begin{equation}\label{equ.ob2}
	\tilde{y}_{m,g_j}^{\left(t\right)} = \frac{{y}_{m,g_j}^{\left(t\right)}}{s_{g_j}^{\left(t\right)}} = \left({h}_{m,g_j}^{\left(t\right)}\right)^{*} + \tilde{n}_{m,g_j}^{\left(t\right)},
\end{equation}
where $\tilde{n}_{m,g_j}^{\left(t\right)} = {{n}_{m,g_j}^{\left(t\right)}}/{s_{g_j}^{\left(t\right)}}$ is the equivalent noise. Similar to (\ref{ch_dd}), the space-frequency domain channel $h_{m,g}^{\left(t\right)}$ can be equivalently rewritten as follows
\begin{align}\label{ch_ad}
	\begin{split}	
		h_{m,g}^{\left(t\right)}\triangleq\sum_{i=1}^{L_u}\dot{x}_{g,i}^{\left(t\right)}e^{-j\frac{2\pi}{\lambda}\bm{k}_{{\rm{tx}},i}^{T}\bm{p}_{m}}.
	\end{split}
\end{align}
Here, $\dot{x}_{g,i}^{\left(t\right)} \triangleq \tilde{x}_{i} {f}_{\rm{rx}}\left(\vartheta_{i},\varphi_i\right) {f}_{\rm{tx}}^{\left(t\right)}\left(\theta_{i},\phi_i\right) e^{-j\frac{2\pi }{\lambda}\bm{k}_{{\rm{rx}},i}^{T}\bm{q}}\cdot
e^{-j2\pi\tau_{i}f_{g}}$ is independent of the antenna index $m$ since we assume that the same radiation pattern ${f}_{\rm{tx}}^{\left(t\right)}\left(\theta,\phi\right)$ is adopted for all antennas at the BS at a given time $t$. 

We introduce a subspace-based method, which yields super-resolution estimates of the AoAs. Based on the observation model (\ref{equ.ob2}) and channel model (\ref{ch_ad}), the received signal can be rewritten as 
\begin{equation}\label{rg_3}
	\tilde{\bm{y}}_{g_j}^{\left(t\right)} = \bm{E}_a\check{\bm{x}}_{g_j}^{\left(t\right)} + \tilde{\bm{n}}_{g_j}^{\left(t\right)},
\end{equation}
where $\tilde{\bm{y}}_{g_j}^{\left(t\right)} = \left[\tilde{y}_{1,g_j}^{\left(t\right)}, \tilde{y}_{2,g_j}^{\left(t\right)}, \ldots, \tilde{y}_{M,g_j}^{\left(t\right)}\right]^{T}\in \mathbb{C}^{M}$ and $\tilde{\bm{n}}_{g_j}^{\left(t\right)} = \left[\tilde{n}_{1,g_j}^{\left(t\right)}, \tilde{n}_{2,g_j}^{\left(t\right)}, \ldots, \tilde{n}_{M,g_j}^{\left(t\right)}\right]^{T}\in \mathbb{C}^{M}$ are the aggregated observations and noise samples, respectively;  $\check{\bm{x}}_{g_j}^{\left(t\right)} = \left[\left(\dot{x}_{g_j,1}^{\left(t\right)}\right)^{*}, \left(\dot{x}_{g_j,2}^{\left(t\right)}\right)^{*},\ldots, \left(\dot{x}_{g_j,L_u}^{\left(t\right)}\right)^{*} \right]^{T} \in \mathbb{C}^{L_u}$ contains the equivalent spatial domain multipath channel gains, and $\bm{E}_a = \left[\bm{e}_a^{1}, \bm{e}_a^{2},\ldots, \bm{e}_a^{L_u}\right] \in \mathbb{C}^{M\times L_u}$ with $\bm{e}_a^{i} = \left[e^{j\frac{2\pi}{\lambda}\bm{k}_{{\rm{tx}},i}^{T}\bm{p}_1}, e^{j\frac{2\pi}{\lambda}\bm{k}_{{\rm{tx}},i}^{T}\bm{p}_2}, \ldots, e^{j\frac{2\pi}{\lambda}\bm{k}_{{\rm{tx}},i}^{T}\bm{p}_M} \right]^{T} \in \mathbb{C}^{M}, \forall i \in \left\{1,2, \ldots, L_u\right\}$.
By aggregating $JT$ snapshots of the observations into columns, we have 
\begin{equation}
	\tilde{\bm{Y}}_a =\bm{{E}}_a\bm{X}_a + \tilde{\bm{N}}_a,
\end{equation}
where $\tilde{\bm{Y}}_a = \left[\tilde{\bm{y}}_{g_1}^{\left(1\right)},\ldots,\tilde{\bm{y}}_{g_J}^{\left(T\right)}\right] \in \mathbb{C}^{M\times JT}$, $\bm{X}_a = \left[\check{\bm{x}}_{g_1}^{\left(1\right)},\ldots,\check{\bm{x}}_{g_J}^{\left(T\right)}\right]\in \mathbb{C}^{L_u \times JT}$, and $\tilde{\bm{N}}_a = \left[\tilde{\bm{n}}_{g_1}^{\left(1\right)}, \ldots, \tilde{\bm{n}}_{g_J}^{\left(T\right)}\right] \in \mathbb{C}^{M\times JT}$.

Furthermore, recalling the system model in Section~\ref{S2}, we have $\bm{k}_{{\rm{tx}},i}= \begin{bmatrix}\sin\theta_{i}\cos\phi_i,\sin\theta_i\sin\phi_{i},\cos\theta_{i}\end{bmatrix}^{T}$ and $\bm{p}_m = \left[0, \frac{2m_y+1-M_y}{2} d, \frac{M_z-2m_z-1}{2} d\right]$. Assuming antenna spacing $d = \frac{\lambda}{2}$, the phase term, $\delta_{i,m} \triangleq \frac{2\pi}{\lambda}\bm{k}^{T}_{{\rm{tx}},i}\bm{p}_m, \forall m$, for each element of $\bm{e}_a^{i}$ can be written as follows
\begin{align}
	\delta_{i,m} =
	m_y\mu_{i} + m_z\nu_{i} + \tilde{\delta}_i,
\end{align}
where $\mu_{i} \triangleq \pi  \sin\theta_{i}\sin\phi_{i} $, $\nu_{i} \triangleq -\pi\cos\theta_{i}$, and $\tilde{\delta}_i \triangleq \pi  \sin\theta_{i}\sin\phi_{i} \frac{1-M_y}{2}  + \left(-\pi\cos\theta_{i}\right)\frac{1-M_z}{2}$ do not depend on the antenna index. Therefore, the observations model can be rewritten as 
\begin{equation}\label{ESPRIT-1}
	\tilde{\bm{Y}}_a = \tilde{\bm{E}}_a\tilde{\bm{X}}_a + \tilde{\bm{N}}_a,
\end{equation}
where $\tilde{\bm{E}}_a = \bm{C}_{y}\circledast \bm{C}_{z}$ and $\bm{\tilde{X}}_a = \bm{\Phi}\bm{X}_a$ with the following definitions,
\begin{subequations}\label{ESPRIT-2}
	\begin{align}
		\bm{C}_{y} &= \left[\tilde{\bm{\zeta}}(\mu_1),\tilde{\bm{\zeta}}(\mu_2),\ldots,\tilde{\bm{\zeta}}(\mu_{L_p})\right]\in \mathbb{C}^{M_y\times L_u},\\
		\bm{C}_{z} &= \left[\tilde{\bm{\zeta}}(\nu_1),\tilde{\bm{\zeta}}(\nu_2),\ldots,\tilde{\bm{\zeta}}(\nu_{L_u})\right]\in \mathbb{C}^{M_z\times L_u},\\
		\tilde{\bm{\zeta}}\left(\mu_{i}\right) &= \left[1, e^{j\mu_{i}},\ldots,e^{j(M_y-1)\mu_{i}}\right]^{T}\in \mathbb{C}^{M_y}, \\
		\tilde{\bm{\zeta}}\left(\nu_{i}\right) &= \left[1, e^{j\nu_{i}},\ldots,e^{j(M_z-1)\nu_{i}}\right]^{T}\in \mathbb{C}^{M_z}
	\end{align}
\end{subequations}
\begin{equation}\label{ESPRIT-3}
	\!\!\!\!\!\!\!\!\!\!\!\!
	\bm{\Phi} = \text{diag}\left(\left[\tilde{\delta}_1,\tilde{\delta}_2,\ldots,\tilde{\delta}_{L_u}\right]\right) \in \mathbb{C}^{L_u\times L_u}.
\end{equation}
Here, $\tilde{\bm{\zeta}} \left(\cdot\right)$ is an intermediate variable. Based on (\ref{ESPRIT-1})-(\ref{ESPRIT-3}), the estimation problem of $\left\{\mu_{i}\right\}_{i=1}^{L_u}$ and $\left\{\nu_{i}\right\}_{i=1}^{L_u}$ can be efficiently solved by the 2D-ESPRIT algorithm in \cite{Liao_Tcom}. 
Based on the estimates $\left\{\hat{\mu}_{i}\right\}_{i=1}^{L_p}$ and $\left\{\hat{\nu}_{i}\right\}_{i=1}^{L_p}$, the received zenith and azimuth angles can be respectively obtained as
	\begin{align}\label{equ.estAng}
		\hat{\theta}_{i} = \arccos\left(-\frac{\hat{\nu}_{i}}{\pi}\right),
		\hat{\phi}_{i} = \arcsin\left(\frac{\hat{\mu}_{i}}{\sqrt{\pi^2-\hat{\nu}_{i}^2}}\right), \forall i.
	\end{align}
Then, we can reconstruct the matrices $\hat{\bm{B}}_m$ and $\hat{\bm{\mathit\Omega}}$ using $\left[\hat{\bm{B}}_{m}\right]_{i,i} = e^{-j\left(m_y\hat{\mu}_{i} + m_z\hat{\nu}_{i} + \tilde{\delta}_i\right)}, \forall i$ and $\left[\hat{\bm{\mathit\Omega}}\right]_{i,k} = \omega_{k}\left(\hat{\theta}_{i},\hat{\phi}_{i}\right), \forall i,k$, respectively. 

Additionally, since $\left(\hat{{h}}_{m,g}^{\left(t\right)}\right)^{*} $ has been estimated in the sCSI estimation stage, as described in Section \ref{Sec-sCSI}, we can construct the corresponding observation model based on the known relationship between the sCSI and eCSI, namely, $\left(h_{m,g}^{\left(t\right)}\right)^{*} = \left(\bm{\alpha}^{\left(t\right)}\right)^{T}\bm{q}_{m,g} = \left(\bm{\alpha}^{\left(t\right)}\right)^{T}\bm{\mathit{\Omega}}^{H}\bm{B}_{m}^{H} \bm{r}_g$, where $\bm{r}_g = \left( \bm{f}_{\rm{rx}}^{T}\bm{A}\bm{\mathit\Sigma}_g\right)^{H}$ is the equivalent channel gain. 
Again, we emphasize that $\bm{\alpha}^{\left(t\right)}$ is independent of the antenna index $m$ but depends on time index $t$. 
This is because we assume that all the BS antennas adopt the same radiation pattern at a given time $t$ and the antenna pattern is altered across time. 
By stacking $MT$ estimates collected by different antennas and for different pilot symbols, we have
\begin{equation}\label{equ_ls_r}
	\hat{\bm{h}}_{g}^{*} = \tilde{\bm{\mathit\Upsilon}}\bm{r}_g, 
\end{equation}
where $\hat{\bm{h}}_{g}^{*} =  \left[\left(\hat{{h}}_{1,g}^{\left(1\right)}\right)^{*}, \ldots, \left(\hat{{h}}_{M,g}^{\left(T\right)}\right)^{*}\right]^{T}\in \mathbb{C}^{MT}$ and $\tilde{\bm{\mathit\Upsilon}} = \left[\hat{\bm{B}}_{1}\hat{\bm{{\mathit{\Omega}}}}\bm{\alpha}^{\left(1\right)},\ldots, \hat{\bm{B}}_{M}\hat{\bm{{\mathit{\Omega}}}}\bm{\alpha}^{\left(T\right)} \right]^{H} \in \mathbb{C}^{MT\times L_u}$. Therefore, if the condition $MT > L_u$ is satisfied, based on (\ref{equ_ls_r}), the equivalent channel gain vector on the $g$-th subcarrier can be estimated using the LS method, i.e.,
\begin{equation}\label{rg_1}
	\hat{\bm{r}}_g  = \left(\tilde{\bm{\mathit\Upsilon}}^{H}\tilde{\bm{\mathit\Upsilon}}\right)^{-1}\tilde{\bm{\mathit\Upsilon}}^{H}\hat{\bm{h}}_g^{*}, \forall g.
\end{equation}
Finally, given the estimated channel gain $\hat{\bm{r}}_g$ and matrices $\hat{\bm{B}}_m$, and $\hat{\bm{\mathit{\Omega}}}$, the eCSI can be reconstructed as follows
\begin{equation}\label{rg_2}
	\hat{\bm{q}}_{m,g}^{H} = \hat{\bm{r}}_g^{H}\hat{\bm{B}}_m\hat{\bm{\mathit{\Omega}}}, \forall m,g.
\end{equation}
Since this method is based on the 2D-AoA estimation using the ESPRIT algorithm in the angular domain, we refer to it as the 2D-ESPRIT algorithm. The overall sCSI and eCSI estimation algorithm is summarized in Algorithm \ref{alg2}.

\begin{table*}[!t]
	\centering
	\captionsetup{name={TABLE}, font = {footnotesize}, justification = raggedright,labelsep=period}
	\caption{Computational complexity of Algorithm \ref{alg2}}
	\label{Tb.2}
	\resizebox{0.9\textwidth}{!}{
		\renewcommand\arraystretch{2}{
			\begin{tabular}{|cc|c|l|}
				\hline
				\multicolumn{2}{|c|}{\textbf{Steps}}                                                                                                  & \textbf{Operations} & \multicolumn{1}{c|}{\textbf{Complexity Order/UE}}                                                    \\ \hline
				\multicolumn{2}{|c|}{\multirow{2}{*}{\textbf{Step 1} (sCSI estimation)}}                                                              & \multirow{2}{*}{1D-ESPRIT}           & \multirow{2}{*}{$J^3+2J^2MT+4JL_u^2+2L_u^3$}        \\
				\multicolumn{2}{|c|}{}                                          &                                      &                                                                                       \\ \hline
				\multicolumn{1}{|c|}{\multirow{3}{*}{\textbf{Step 2} (eCSI estimation)}} & \multirow{2}{*}{\textbf{Step 2 (a)} (AoAs estimation)}              & \multirow{2}{*}{2D-ESPRIT}           & \multirow{2}{*}{$M^3+2M^2JT+4ML_u^2+2L_u^3$} \\
				\multicolumn{1}{|c|}{}                                          &                                                            &                                      &                                                                                       \\ \cline{2-4} 
				\multicolumn{1}{|c|}{}                                          & \textbf{Step 2 (b)} (eCSI reconstruction) & Lines 11-12                          & $GMTL_u + GMKL_u + 2KMTL_u + 2MTL_u^2+\frac{2}{3}L_u^3$                                            \\ \hline
				\multicolumn{3}{|c|}{Overall Complexity / UE}                    &  $M^3+2M^2JT+2J^2MT+J^3+GMKL_u+\left(4J+4M+2MT\right)L_u^2+\frac{14}{3}L_u^3$                                                                                \\ \hline
	\end{tabular}}}
	\vspace{-3mm}
\end{table*}

\begin{algorithm}[!t]
	\small
	\KwIn{Uplink received signal $\left\{y_{m,g_j}^{\left(t\right)}\right\}_{m=1, j=1, t=1}^{M,J,T}$, radiation pattern coefficient vector at different pilot symbols $\left\{\bm{\alpha}_{m}^{\left(t\right)}\right\}_{t=1,m=1}^{T,M}$, pilot sequence $\left\{s_{g_j}^{\left(t\right)}\right\}_{j=1,t=1}^{J,T}$.} 
	\KwOut{sCSI $\hat{h}_{m,g}^{\left(t\right)}, \forall m,g,t,$ and eCSI $\hat{\bm{q}}_{m,g}, \forall m,g $.}
	\% \textbf{Step 1: Estimate full-band sCSI } \\
	Estimate  $\left\{\hat{\kappa}_{i}\right\}_{i=1}^{L_u}$by applying 1D-ESPRIT to (\ref{ce:sCSI4});\\
	Estimate multipath delays $\left\{\hat{\tau}_{i}\right\}_{i=1}^{L_u}$and equivalent channel gains $\left\{\hat{x}_{m,i}^{\left(t\right)}\right\}_{i=1}^{L_u}$ using (\ref{ce:tau}) and (\ref{ce:ls1}), respectively;\\
	Reconstruct sCSI $\hat{h}_{m,g}^{\left(t\right)}, \forall m,g,t$, according to $\hat{h}_{m,g}^{\left(t\right)} = \sum_{i=1}^{L_u}\hat{x}_{m,i}^{\left(t\right)}e^{-j2\pi\hat{\tau}_{i}f_{g}}, \forall m,g,t$;\\
	
	\% \textbf{Step 2: Estimate eCSI } \\
	\% \textbf{Step 2(a): Estimate multipath AoAs} \\
	Estimate $\left\{\hat{\mu_{i}}\right\}_{i=1}^{L_u}$ and $\left\{\hat{\nu_{i}}\right\}_{i=1}^{L_u}$ by applying 2D-ESPRIT algorithm to (\ref{ESPRIT-1});\\
	Acquire the zenith angles $\left\{\hat{\theta}_{i}\right\}_{i=1}^{L_u}$ and azimuth angles $\left\{\hat{\phi}_i\right\}_{i=1}^{L_u}$ using (\ref{equ.estAng});\\

	\% \textbf{Step 2(b): Reconstruct the eCSI} \\
	Construct $\hat{\bm{\mathit{\Omega}}}$ and $\hat{\bm{B}}_m$ using the estimated zenith and azimuth angles;\\
	Estimate the equivalent channel gain $\left\{\bm{r}_g\right\}_{g=1}^{G}$ using (\ref{rg_1});\\
	Reconstruct the eCSI $\hat{\bm{q}}_{m,g}, \forall m,g,$ using (\ref{rg_2}); \\ 
	\Return $\left\{\hat{h}_{m,g}^{\left(t\right)}\right\}_{m=1,g=1,t=1}^{M,G,T}$ and $\left\{\hat{\bm{q}}_{m,g}\right\}_{m=1,g=1}^{M,G}$.
	\caption{sCSI and eCSI Estimation Algorithm for Each UE}
	\label{alg2} 
	\LinesNumbered	
\end{algorithm}

\subsection{Complexity Analysis}
The computational complexity of Algorithm \ref{alg2} is mainly attributed to the 1D-ESPRIT operation in line 2, the 2D-ESPRIT operation in line 7, and the eCSI reconstruction in lines 11-12. Following the complexity analysis presented in \cite{Liao_Tcom}, we obtain the computational complexities for 1D-ESPRIT and 2D-ESPRIT, which are approximately  $\mathcal{O}\left(J^3+2J^2MT+4JL_u^2+2L_u^3\right)$ and $\mathcal{O}\left(M^3+2M^2JT+4ML_u^2+2L_u^3\right)$, respectively. Neglecting terms significantly smaller than others, we summarize the overall channel estimation complexity per UE in Table \ref{Tb.2}.

\begin{figure*}[!t]
	\subfigure[]{
		\begin{minipage}[t]{0.33\linewidth}
			\centering
			\includegraphics[width=\textwidth]{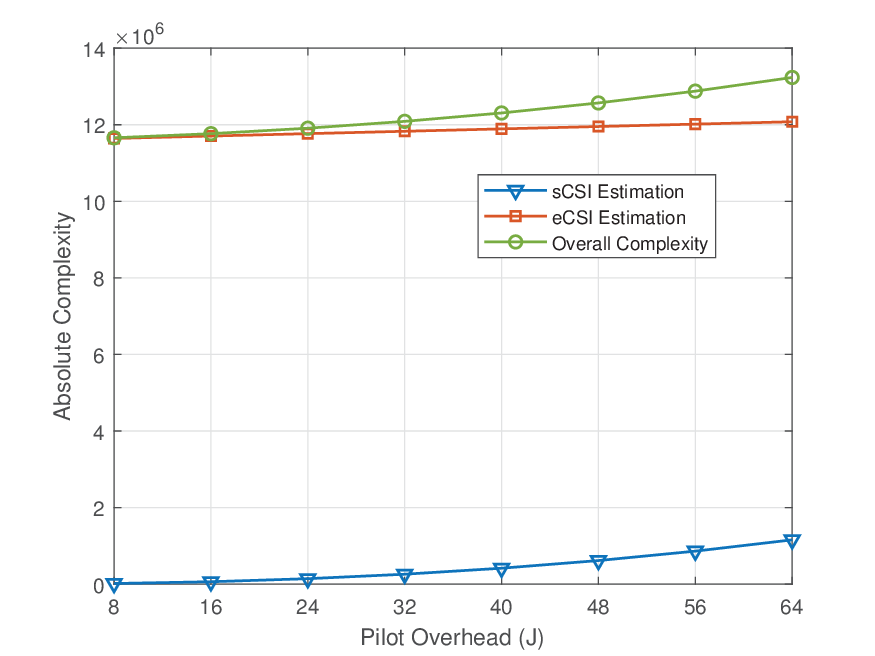}
		\end{minipage}
	}%
	\subfigure[]{
		\begin{minipage}[t]{0.33\linewidth}
			\centering
			\includegraphics[width=\textwidth]{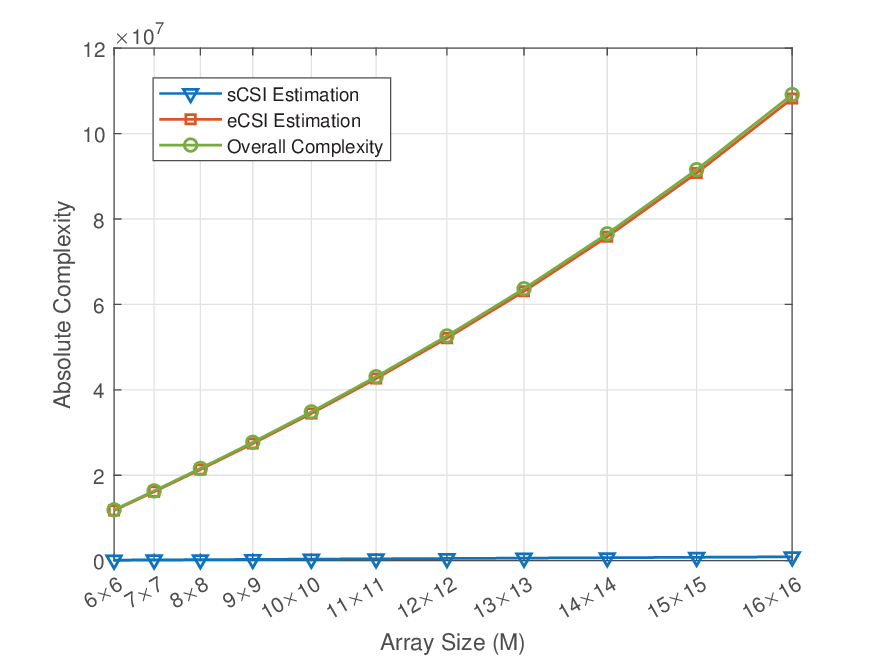}
		\end{minipage}
	}%
	\subfigure[]{
		\begin{minipage}[t]{0.33\linewidth}
			\centering
			\includegraphics[width=\textwidth]{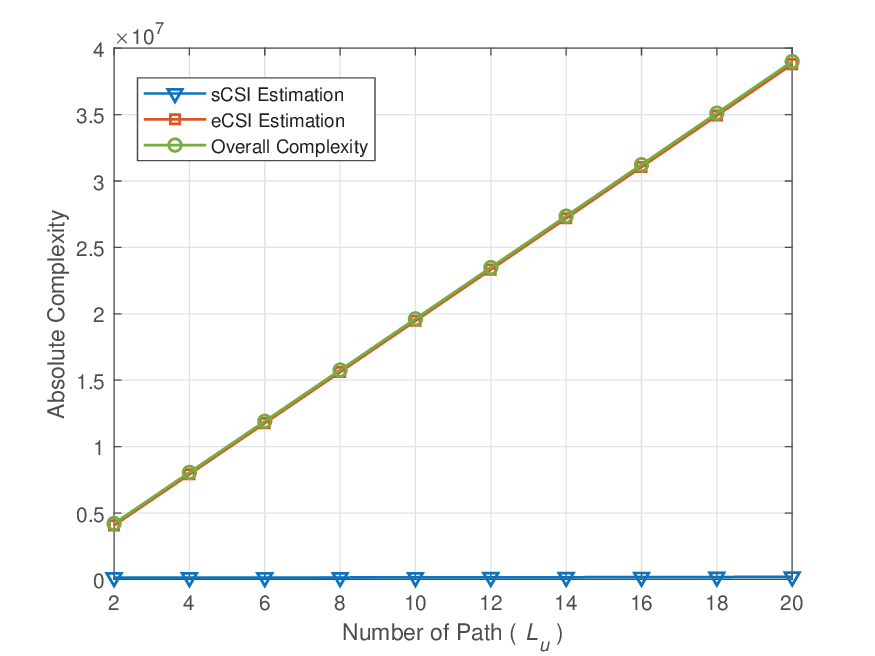}
		\end{minipage}
	}%
	\centering
	\vspace*{-3mm} 
	\captionsetup{font={footnotesize}, singlelinecheck = off,name={Fig.}, labelsep=period}
	\caption{a) Complexity versus the pilot overhead $J$, b) complexity versus the array size $M$, c) complexity versus the multipath number $L_u$.}
	\label{fig_complexity_ce}
	\vspace*{-5mm}
\end{figure*}

In Figs. \ref{fig_complexity_ce} (a)-(c), we illustrate the number of complex multiplications of the proposed channel estimation algorithm as a function of the key algorithm parameters. While our theoretical analysis predicted that the overall complexity includes cubic terms of the frequency-domain pilot overhead $J$, number of antennas $M$, and number of multipath components $L_u$, the simulation results indicate that the number of antennas (Fig. \ref{fig_complexity_ce} (b)) and number of multipaths (Fig. \ref{fig_complexity_ce} (c)) primarily dominate the computational complexity. Conversely, variations in the pilot overhead $J$ do not markedly increase complexity. Given the relatively small coefficients of the cubic terms for $J, M$, and $L_u$, for the considered parameter range, the overall complexity exhibits an approximate quadratic growth with respect to the pilot overhead and array size, and a linear growth with respect to the number of multipath components. Furthermore, we observe that the eCSI estimation step accounts for a significant part of the overall complexity of the algorithm.

\section{Simulation Results}\label{S6}
This section evaluates the performance of the proposed RmMIMO architecture and compares it with that of the fully digital TmMIMO architecture. The evaluation focuses on the single-cell scenario, as illustrated in Fig. \ref{Fig.scenario}. The mMIMO BS is equipped with a UPA comprising pattern-reconfigurable antennas, while each UE is equipped with a single omnidirectional antenna. Unless otherwise specified, the BS uses a $6\times 6$ array. In its local coordinate system, the BS array is located in the $zy$ plane, with its center at the origin. The simulated multipath channel is generated following the methodology described in \cite{38.901}. The number of paths is assumed to be $ L_u = 6$. The azimuth AoD for each path at the BS is assumed to be uniformly distributed within  $\mathcal{U}[-60^{\circ},60^{\circ}]$, while the zenith AoD for each path follows $\mathcal{U}[60^{\circ},150^{\circ}]$. Furthermore, we assume a wideband system with $G = 512$ subcarriers and a subcarrier spacing of $\Delta f = 30$ kHz. The maximum delay spread is assumed to be $\tau_{\max} = 100$ ns. For both TmMIMO and RmMIMO, we adopt ZF digital domain precoding. 

\subsection{Performance of Channel Estimation}
In this subsection, we present simulation results for the proposed sCSI and eCSI estimation methods. We evaluate the performance using the NMSE as the metric. The estimation errors for sCSI and eCSI are respectively defined as
\begin{align}
\text{NMSE-S} &= \mathbb{E}\left\{ \frac{\sum_{m=1,g=1,t=1}^{M,G,T}|{h}_{m,g,t}-\hat{h}_{m,g,t}|^2}{\sum_{m=1,g=1,t=1}^{M,G,T}|{h}_{m,g,t}|^2}\right\}, \\
\text{NMSE-E} &= \mathbb{E}\left\{ \frac{\sum_{m=1,g=1}^{M,G}\|\bm{q}_{m,g}-\hat{\bm{q}}_{m,g}\|^2}{\sum_{m=1,g=1}^{M,G}\|\bm{q}_{m,g}\|^2}\right\}.
\end{align}
Additionally, the received signal-to-noise ratio (SNR) for uplink channel estimation is defined as $\text{SNR}_{u} = \frac{\mathbb{E}\left\{|{h}_{m,g_l,t}^{*} s_{g_l,t}|^{2}\right\}}{\mathbb{E}\left\{|{n}_{m,g_l,t}|^2\right\}}$. 

To demonstrate the effectiveness of our proposed ESPRIT-based channel estimation methods, we compare them with conventional simultaneous orthogonal matching pursuit (OMP) -based schemes \cite{SOMP}.
	
$\bullet$ \textbf{Benchmark for sCSI estimation:} By utilizing the delay domain sparsity of the channel, the full-band sCSI recovery can be modeled as a compressed sensing problem. We adopt a random pilot allocation method in the frequency domain, and an oversampled DFT dictionary is adopted for improved sparsity \cite{wzw_tvt}. For simplicity, we refer to this approach as delay domain-OMP (DD-OMP).

$\bullet$ \textbf{Benchmark for eCSI estimation:} OMP can be used for AoA estimation by utilizing the angular domain sparsity of the channel. First, an angular domain oversampled DFT dictionary is constructed for improved AoA estimation. Then, the LS method is used for channel gain estimation and eCSI reconstruction, as in (\ref{equ_ls_r})-(\ref{rg_2}). For simplicity, we refer to this approach as angular domain-OMP (AD-OMP).

\begin{figure}[!t]
	\begin{center}
		\includegraphics[width = 1\columnwidth]{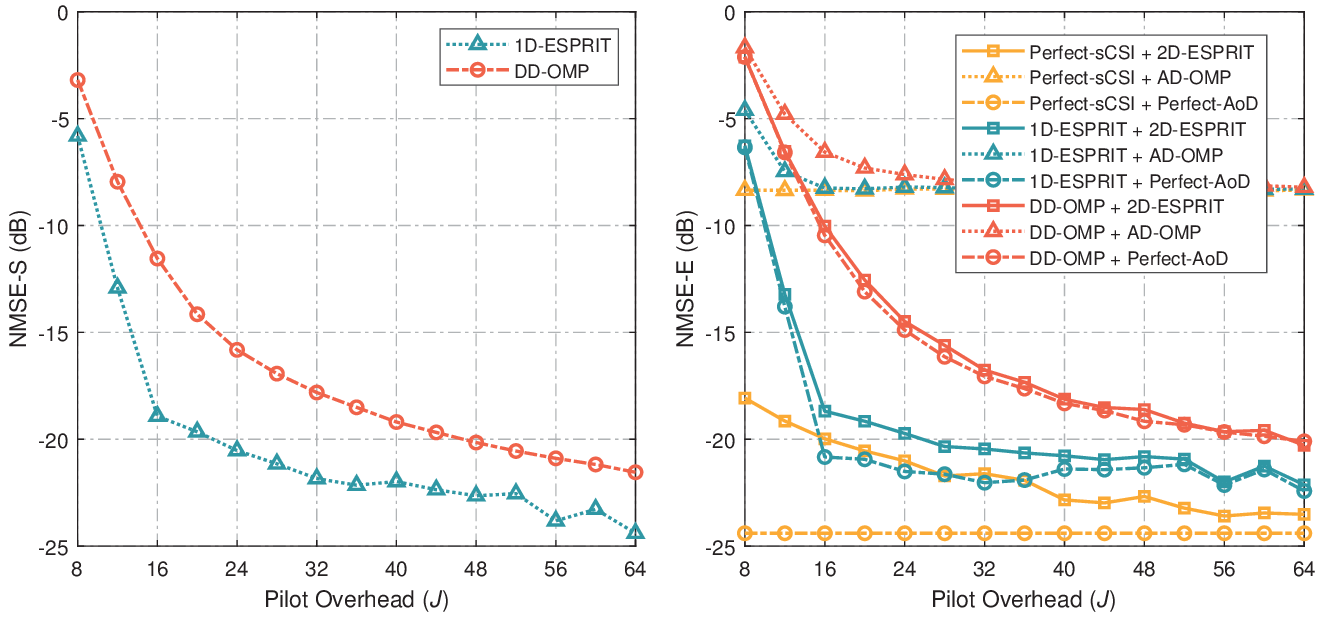}
	\end{center}
	\vspace*{-3mm}
	\captionsetup{font={footnotesize}, singlelinecheck = off,name={Fig.}, labelsep=period}
	\caption{NMSE versus frequency domain pilot overhead $J$ for different channel estimation methods, for $M = 6\times 6$, $K = 100$, $\text{SNR}_u = 15$ dB, and $T = 3$.}
	\label{fig.ce_sim1} 
	\vspace*{-3mm}
\end{figure}

Figure~\ref{fig.ce_sim1} investigates the performance of different channel estimation algorithms in terms of	
$\text{NMSE-S}$ and $\text{NMSE-E}$ as a function of the frequency domain pilot overhead $J$. The simulation setup assumes a BS with a UPA comprising $M = 6\times 6$ antennas. $\text{SNR}_{u}$ is set to $15$ dB, the number of pilot symbols in the time domain is $T = 3$, and the number of SH basis functions is $K = 10^2 = 100$, i.e., the radiation patterns are expanded over $100$ coefficients in the EM domain. From Fig. \ref{fig.ce_sim1}, we can make the following observations: 
For the conventional sCSI estimation, the performance of both the 1D-ESPRIT and DD-OMP algorithms improves as the pilot overhead in the frequency domain increases. Specifically, 1D-ESPRIT outperforms DD-OMP due to its high-resolution estimation capability of multipath delays for a sufficiently large bandwidth. For the eCSI estimation problem, we consider three types of sCSI information as input: Perfect-sCSI, sCSI estimated by 1D-ESPRIT, and sCSI estimated by DD-OMP. We compare three eCSI estimation methods: 2D-ESPRIT, AD-OMP, and Perfect-AoD-based eCSI estimation. Here, Perfect-AoD indicates the use of perfect AoD information in the eCSI reconstruction process (\ref{rg_1}), (\ref{rg_2}). Perfect-sCSI input yields an NMSE-E lower bound for all eCSI estimation methods. Due to the undersampled frequency domain observations and the receiving noise in the channel estimation, this bound is not achievable.
Furthermore, the use of more accurate AoD information (from AD-OMP to 2D-ESPRIT to Perfect-AoD) improves the  $\text{NMSE-E}$ performance.  
Additionally, for a given eCSI estimation method, as $J$ increases, the DD-OMP-based approach achieves asymptotically similar $\text{NMSE-E}$ performance as 1D-ESPRIT. This occurs as the accuracy of sCSI estimation improves for larger values of $J$. Overall, the proposed ESPRIT-based schemes attain the desired target NMSE with lower pilot overhead compared to the OMP-based schemes for both sCSI and eCSI estimation. As outlined in Algorithm \ref{alg2}, obtaining eCSI information based on estimated sCSI does not incur additional cost, resulting in a comparable pilot overhead as TmMIMO systems.

\begin{figure}[!t]
	\begin{center}
		\includegraphics[width = 1\columnwidth]{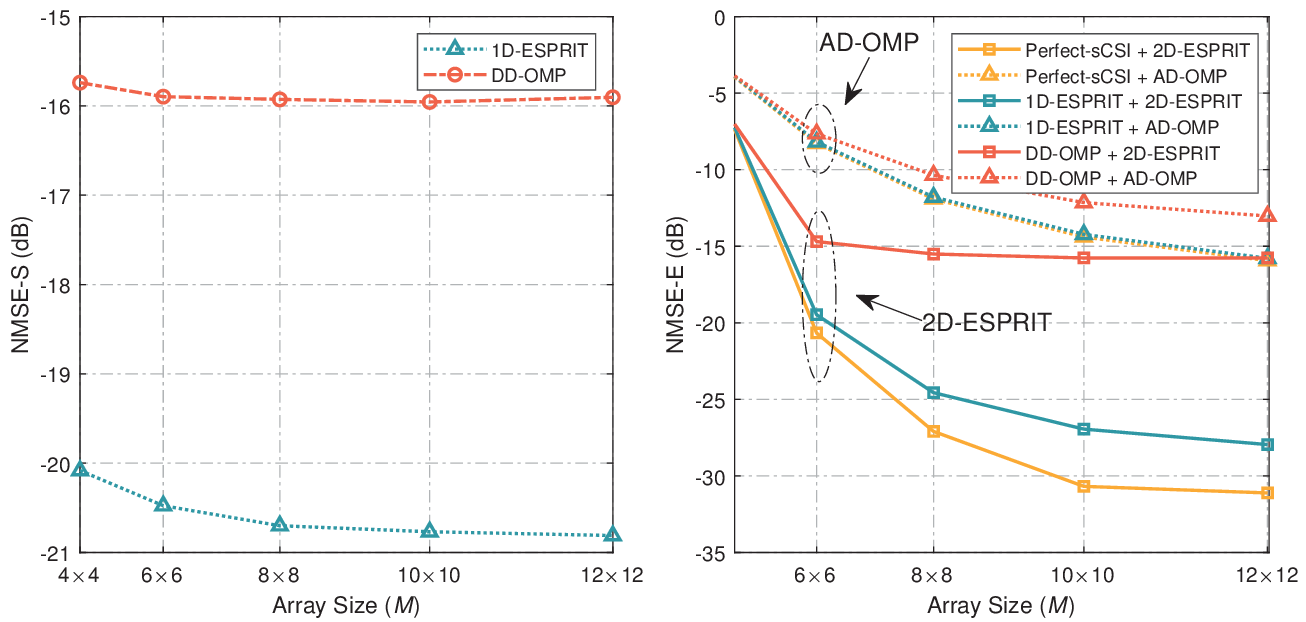}
	\end{center}
	\vspace*{-3mm}
	\captionsetup{font={footnotesize}, singlelinecheck = off,name={Fig.}, labelsep=period}
	\caption{NMSE versus the array size for different channel estimation algorithms when $K = 100$, $\text{SNR}_u = 15$ dB, $J = 24$, and $T = 3$.}
	\label{fig.ce_sim2} 
	\vspace*{-3mm}
\end{figure}
Figure \ref{fig.ce_sim2} investigates the NMSE performance for different channel estimation algorithms, in relation to the size of the BS antenna array. The simulation parameters are set as $K = 100$, $\text{SNR}_u = 15$ dB, $J = 24$, and $T = 3$. Notably, enlarging the receive antenna array size enhances the channel estimation performance for both sCSI and eCSI. The improved sCSI performance is attributed to the increased number of observations of the delay-domain channel when using a larger antenna array, consequently reducing the equivalent noise for DD-OMP and 1D-ESPRIT. As the frequency domain pilot overhead is not large ($J=24$), 1D-ESPRIT demonstrates significantly better NMSE-S performance compared to DD-OMP, as illustrated in Fig. \ref{fig.ce_sim1}. For eCSI estimation, increasing the array size enhances the spatial resolution of the BS, results in a significant improvement in AoA estimation accuracy. Regarding $\text{NMSE-E}$, Perfect-sCSI provides a performance upper bound for the other algorithms. The combination of 1D-ESPRIT and 2D-ESPRIT yields the smallest performance gap compared to the Perfect-sCSI case, demonstrating its good performance for eCSI estimation.

\begin{figure}[!t]
	\begin{center}
		\includegraphics[width = 1\columnwidth]{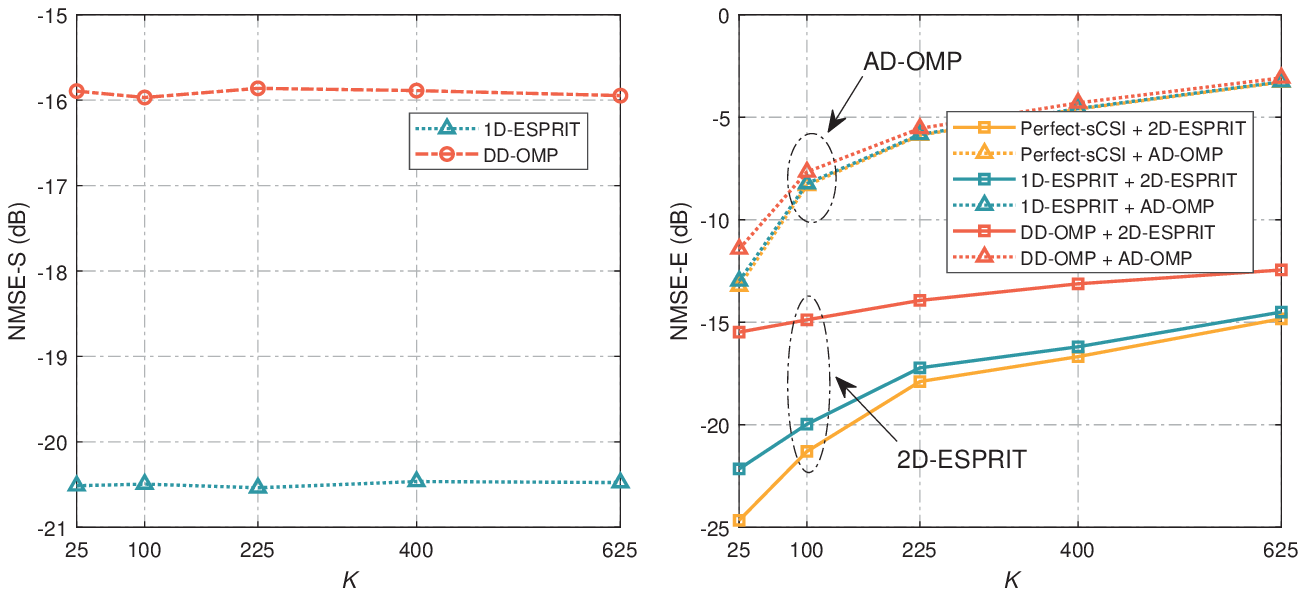}
	\end{center}
	\vspace*{-3mm}
	\captionsetup{font={footnotesize}, singlelinecheck = off,name={Fig.}, labelsep=period}
	\caption{NMSE versus different values of $K$ for different channel estimation algorithms, when $M = 6\times 6$, $\text{SNR}_u = 15$ dB, $J = 24$, and $T = 3$.}
	\label{fig.ce_sim3} 
	\vspace*{-3mm}
\end{figure}

Figure \ref{fig.ce_sim3} shows the NMSE performance as a function of $K$ for different channel estimation schemes, where $M = 6\times 6$, $\text{SNR}_u = 15$ dB, $J = 24$, and $T = 3$. Given the fixed pilot overheads in the frequency domain and time domain, the parameter $K$ serves as an adjustable parameter for the input of the eCSI estimation algorithm, determining the dimension of the estimated eCSI and the corresponding DoF for antenna radiation pattern design.
In Fig. \ref{fig.ce_sim3}, it is observed that the average $\text{NMSE-S}$ remains relatively constant irrespective of the value of $K$, whereas the average $\text{NMSE-E}$ increases as $K$ grows. This behavior can be attributed to the fact that the sCSI estimation process does not involve EM domain operations.
In the context of eCSI estimation, where the eCSI is defined as $\bm{q}_{u,m,g}^{H} \triangleq \bm{r}_{u,g}\bm{B}_{u,m}\bm{\mathit\Omega}_u\in \mathbb{C}^{1\times K}$, errors for high $K$ values primarily stem from inaccuracies in estimating $\bm{\mathit{\Omega}}_{u} \in \mathbb{C}^{L_u\times K}$, which also impacts the accuracy of the subsequent channel gain estimation $\bm{r}_{u,g}$ indirectly. This is because the $\left(i,k\right)$-th element of $\bm{\mathit\Omega}_u \in \mathbb{R}^{L_u \times K}$ is given by $\left[\bm{\mathit\Omega}_u\right]_{i,k} = \omega_{k}\left(\theta_{i,u},\phi_{i,u}\right) = Y_{c}^r(\theta_{i,u}, \phi_{i,u})$, where small errors in estimating the AoD $\left(\theta_{i,u},\phi_{i,u}\right)$ can significantly impact the gain of $\omega_{k}\left(\theta_{i,u},\phi_{i,u}\right)$, especially for high-order SH basis functions with intricate shapes.\footnote{Similar to the properties of Fourier transform bases for one-dimensional signals, for SH functions, lower-order basis functions represent functions with smoother shapes, while higher-order basis functions display more complex and intricate shapes.} Notably, the proposed 2D-ESPRIT algorithm outperforms the AD-OMP algorithm for all considered types of sCSI input, underscoring the effectiveness of the proposed approach.

\begin{figure}[!t]
	\begin{center}
		\includegraphics[width = 1\columnwidth]{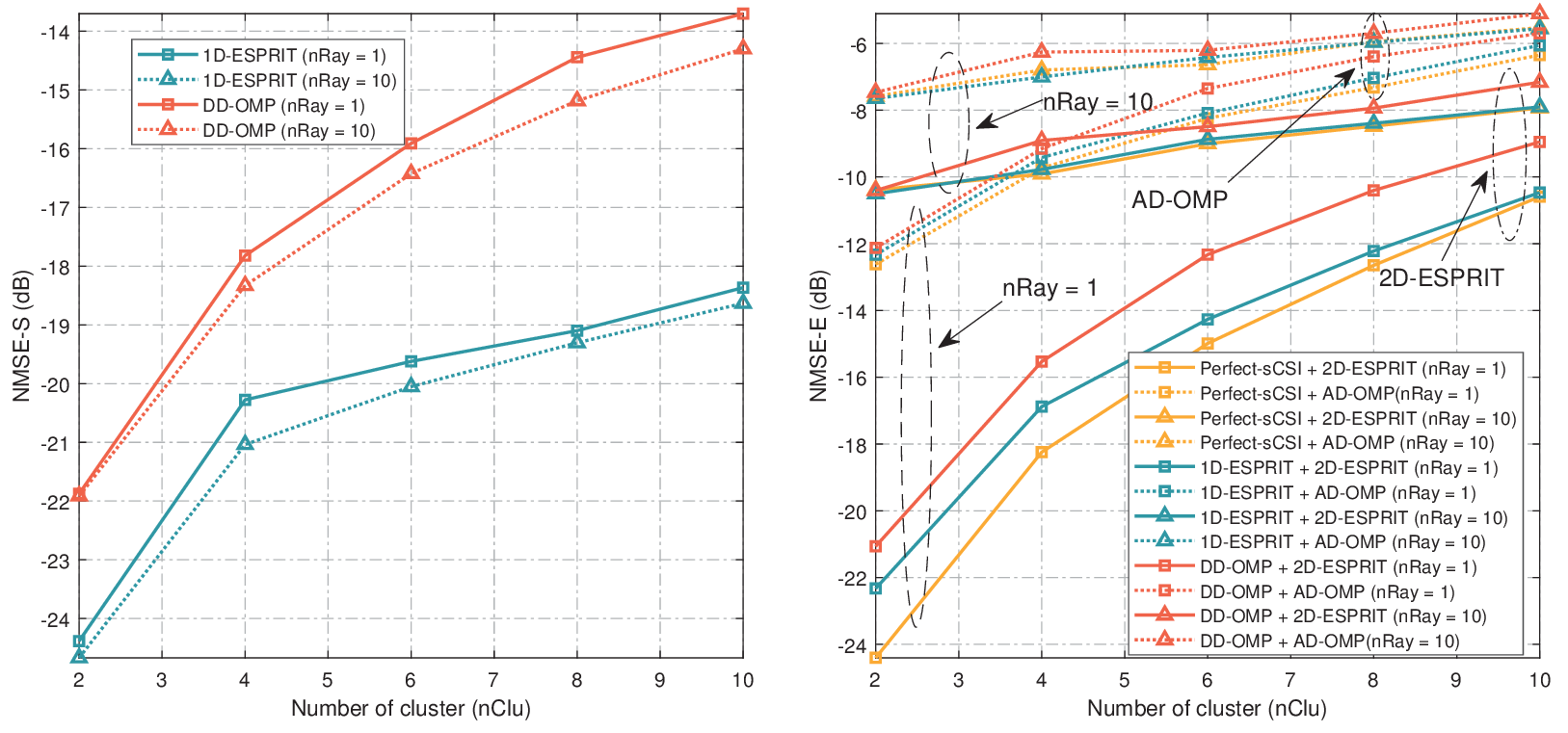}
	\end{center}
	\vspace*{-3mm}
	\captionsetup{font={footnotesize}, singlelinecheck = off,name={Fig.}, labelsep=period}
	\caption{NMSE versus the number of multipath clusters and rays, when $M = 6 \times 6$, $K = 100$, $\text{SNR}_u = 15$ dB, $L = 24$, and $T = 3$.}
	\label{fig.ce_sim4} 
	\vspace*{-3mm}
\end{figure}
Figure \ref{fig.ce_sim4} assesses the impact of the number of multipath components on channel estimation. Additionally, we extend the channel environment to cases where multiple rays exist within each cluster. The simulation parameters are set as follows: $M = 6 \times 6$, $K = 100$, $\text{SNR}_u = 15$ dB, $L = 24$, and $T = 3$. Two cases are considered, namely one ray per cluster and ten rays per cluster. For the latter case, we assume the azimuth angle spread and zenith angle spread at the BS to be $10^{\circ}$ and $3^{\circ}$, respectively. For both sCSI and eCSI estimation, increasing the number of clusters leads to a worse NMSE due to the more complex channel environment that needs to be estimated. For sCSI estimation, increasing the number of rays from one to ten does not deteriorate the NMSE-S severely. This is because, for delay-domain sCSI estimation, different rays within the same cluster approximately share the same delay, according to the channel model in \cite{38.901}. For eCSI estimation, as different rays correspond to different angles in the spatial domain, it becomes more challenging for the BS to distinguish angle differences within the same cluster. Therefore, for both the 2D-ESPRIT and AD-OMP algorithms, increasing the number of rays leads to worse eCSI estimation performance.

\subsection{Performance of EM Domain Precoding}

\subsubsection{Comparison with Benchmarks}
In this section, we evaluate the performance gain achieved by EM domain precoding in multi-user downlink transmission. The SE given in (\ref{eq.se}) is used as the performance metric. Here, we assume that the channel environment is quasi-static and the number of time domain pilot symbols $T$ is negligible compared to the number of data symbols. Therefore, we ignore the SE loss caused by pilot overhead. For TmMIMO, we consider three different baseline antenna patterns, namely the dipole pattern, 3GPP 38.901 pattern \cite{38.901}, and downtilt pattern, as shown in Fig. \ref{Fig.SHOD}. For RmMIMO, we consider both SM and MM optimization.

\begin{figure}[!t]
	\begin{center}
		\includegraphics[width = 1\columnwidth]{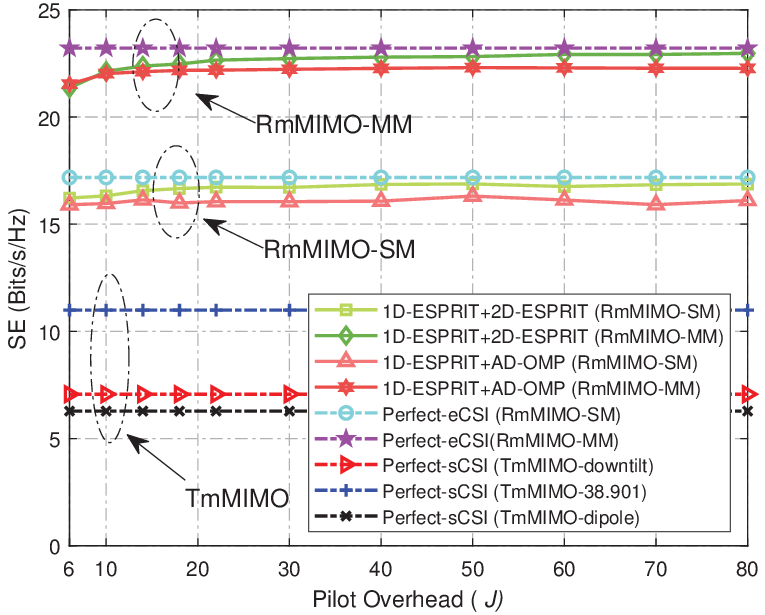}
	\end{center}
	\vspace*{-3mm}
	\captionsetup{font={footnotesize}, singlelinecheck = off,name={Fig.}, labelsep=period}
	\caption{SE versus $J$ for RmMIMO and TmMIMO when $U = 6$, $M = 6\times 6$, $K = 225$, $\text{SNR}_u = 0 \; \text{dB}$, $T = 3$, and $\text{SNR}_d = 15 \; \text{dB}$.}
	\label{fig.pre_sim1} 
	\vspace*{-3mm}
\end{figure}

\begin{figure}[!t]
	\begin{center}
		\includegraphics[width = 1\columnwidth]{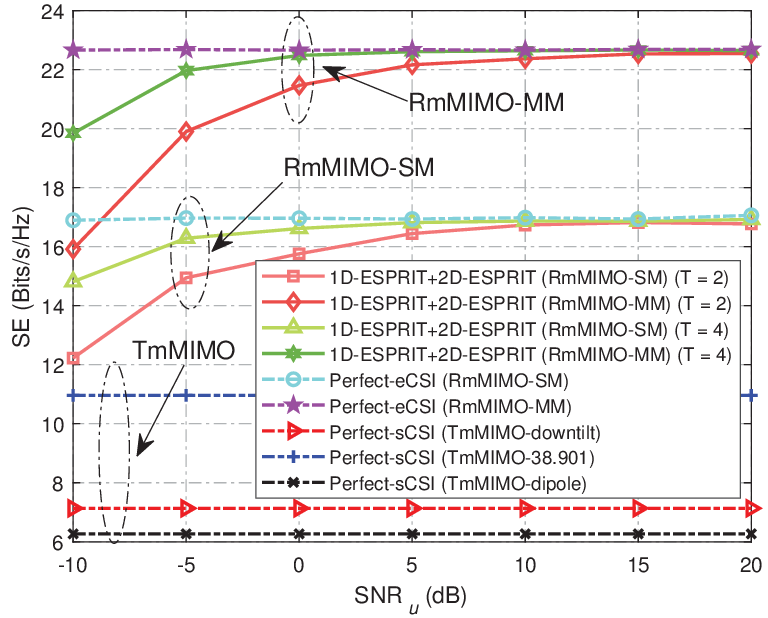}
	\end{center}
	\vspace*{-3mm}
	\captionsetup{font={footnotesize}, singlelinecheck = off,name={Fig.}, labelsep=period}
	\caption{SE versus $\text{SNR}_u$ for RmMIMO and TmMIMO when $U = 6$, $M = 6\times 6$, $K = 225$, $\text{SNR}_{d} = 15$ dB, and $J = 24$.}
	\label{fig.pre_sim2} 
	\vspace*{-3mm}
\end{figure}


Figure \ref{fig.pre_sim1} illustrates the SE gains achieved by RmMIMO over TmMIMO for different numbers of frequency domain pilot symbols $J$. The simulation assumes $U = 6$ UEs are served by a UPA with $M = 6\times 6$ antennas. The received SNR is set to $\text{SNR}_u = 0 $ dB during the uplink channel estimation stage, the number of time domain pilot symbols is $T = 3$, while the received SNR at the downlink transmission stage is set to $\text{SNR}_{d} = \frac{\mathbb{E}\left\{\|\bm{q}_{u,g}^{H}{\bm{\mathit\Lambda}}\bm{W}_{g}\bm{s}_{g}\|_2^{2}\right\}}{\mathbb{E}\left\{\|{n}_{u,g}\|_2^2\right\}} = 15$ dB. Additionally, we assume that the number of SH basis functions is $K = 15^2 = 225$. Notably, since the ESPRIT algorithm requires more observations than the number of paths $L_u=6$, the minimum number of frequency domain pilot symbols is set to $J = 6$. Figure \ref{fig.pre_sim1} reveals that the proposed RmMIMO pattern optimization schemes provide significant SE gains over the TmMIMO design. Specifically, RmMIMO with MM design (RmMIMO-MM) achieves the best performance by customizing the radiation pattern of each antenna. Conversely, RmMIMO with SM design (RmMIMO-SM) provides smaller SE gains over TmMIMO as the antennas are forced to employ the same radiation pattern. TmMIMO systems offer the lowest SE performance due to their fixed radiation pattern. Particularly, the 3GPP 38.901 pattern yields the best performance among the TmMIMO schemes as its radiation pattern aligns with the distribution of the azimuth and zenith AoD. By contrast, the dipole pattern performs poorly as its energy distribution is more uniform in space, leading to energy wastage in unnecessary directions. Furthermore, both RmMIMO-SM and RmMIMO-MM with estimated eCSI approach the performance of Perfect-eCSI as the pilot length $J$ increases, indicating that a higher pilot overhead results in SE gains, as expected. Additionally, the superior accuracy of the eCSI provided by the 2D-ESPRIT method compared to the AD-OMP algorithm results in a reduced SE gap when compared to the Perfect-eCSI scenario.
 
Figure \ref{fig.pre_sim2} examines the SE performance of different precoding schemes as a function of $\text{SNR}_u$. The simulation assumes that $U = 6$, $M = 6 \times 6 $, $K = 225$, $\text{SNR}_{d} = 15$ dB, and $J = 24$. For the RmMIMO system, three different kinds of eCSIs are considered: estimated eCSI for
$T = 2, 4$ time domain pilot symbols and perfect eCSI. For the TmMIMO system, perfect sCSI is utilized for precoding design. For RmMIMO with estimated eCSI, the SE monotonically increases as $\text{SNR}_u$ grows. This is because the eCSI estimation becomes increasingly accurate. Similar to the findings in Fig. \ref{fig.pre_sim1}, RmMIMO-MM achieves the highest SE among the considered schemes. Notably, due to the accurate estimation of eCSI at high $\text{SNR}_u$, both RmMIMO-SM and RmMIMO-MM using the estimated eCSI approach the performance achieved with Perfect-eCSI.

\begin{figure}[!t]
	\begin{center}
		\includegraphics[width = 1\columnwidth]{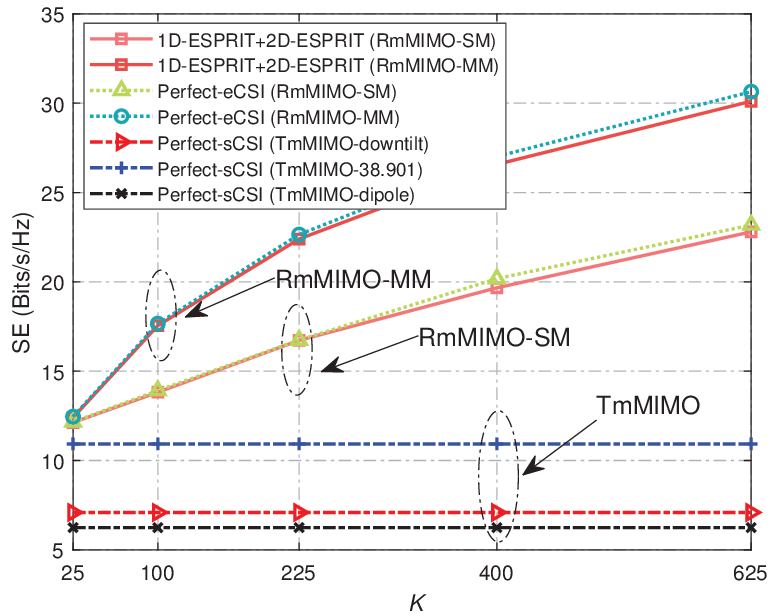}
	\end{center}
	\vspace*{-3mm}
	\captionsetup{font={footnotesize}, singlelinecheck = off,name={Fig.}, labelsep=period}
	\caption{SE versus the value of $K$ for RmMIMO and TmMIMO when $U = 6$, $M = 6\times 6$, $\text{SNR}_{u} = 10$ dB, $J = 24$, $T = 2$, and $\text{SNR}_{d} = 15$ dB.}
	\label{fig.pre_sim3} 
	\vspace*{-3mm}
\end{figure}


Figure \ref{fig.pre_sim3} illustrates the influence of the number of SH basis functions on the SE performance. The simulation assumes $U = 6$,  $M = 6 \times 6 $, $\text{SNR}_{u} = 10$ dB, $J = 24$, $T = 2$, and $\text{SNR}_{d} = 15$ dB. 
For RmMIMO with Perfect-eCSI, a larger value of $K$ provides more design DoF for the radiation pattern functions, thereby leading to an improved SE performance. Conversely, TmMIMO systems rely solely on the sCSI for precoding design, rendering their SE performance unaffected by variations in $K$. For RmMIMO systems utilizing the estimated eCSI obtained via the 1D-ESPRIT+2D-ESPRIT method, a larger value of $K$ increases the dimensionality of the eCSI to be estimated, consequently increasing the estimation error.
Intriguingly, in Fig. \ref{fig.pre_sim3}, the SE for the case of estimated eCSI consistently grows with $K$. This is due to the fact that, for the considered simulation parameters, the channel estimation remains sufficiently accurate to facilitate reliable precoding design, as evidenced in Fig. \ref{fig.ce_sim3}. Consequently, the positive impact of the additional design DoF outweighs the negative effects caused by channel estimation errors, suggesting that a larger $K$ enhances the SE performance overall. Nevertheless, from a more practical point of view, a higher $K$ also leads to a more intricate radiation pattern structure, which presents challenges in practical antenna design and escalates the computational burden for both channel estimation and precoding. 

\begin{figure}[!t]
	\begin{center}
		\includegraphics[width = 1\columnwidth]{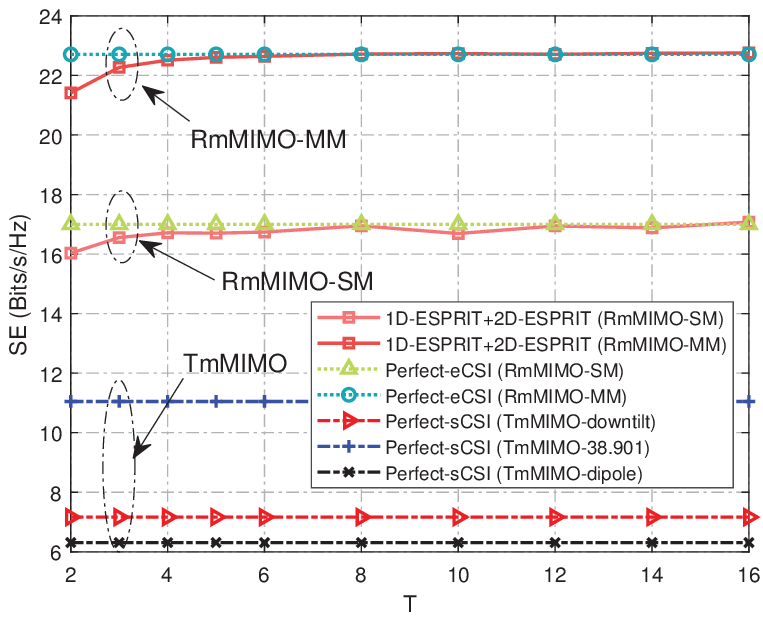}
	\end{center}
	\vspace*{-3mm}
	\captionsetup{font={footnotesize}, singlelinecheck = off,name={Fig.}, labelsep=period}
	\caption{SE versus the number of pilot symbols $T$ in the time domain for RmMIMO and TmMIMO when $U = 6$, $M = 6\times 6$, $K = 225$,  $\text{SNR}_u = 0$ dB, $J = 24$, and $\text{SNR}_d = 15$ dB.}
	\label{fig.pre_sim4} 
	\vspace*{-3mm}
\end{figure}

Figure \ref{fig.pre_sim4} illustrates the SE gains achieved by RmMIMO and TmMIMO as a function of the number of time domain pilot symbols $T$.
The simulation assumes $U = 6$, $M = 6\times 6$, $K=225$, $\text{SNR}_u = 0 \; \text{dB}$, $J = 24$, and $\text{SNR}_d = 15 \; \text{dB}$. 
As depicted in Fig. \ref{fig.pre_sim4}, the SE of RmMIMO systems with estimated eCSI increase with the number of time domain pilot symbols $T$, since the channel estimation performance improves for larger $T$. When $T = 6$, the performance of the proposed schemes can approach that for Perfect-eCSI for both SM and MM. Actually, increasing $T$ mitigates the adverse effects of low uplink $\text{SNR}_u = 0 \; \text{dB}$. On the other hand, when the uplink SNR is large, a low time domain pilot overhead $T$ is sufficient to achieve good performance, as shown in Fig. \ref{fig.pre_sim2}.  

\subsubsection{Illustration of Optimized Radiation Patterns}
\begin{figure}[!t]
	\begin{center}
		\includegraphics[width = 1\columnwidth]{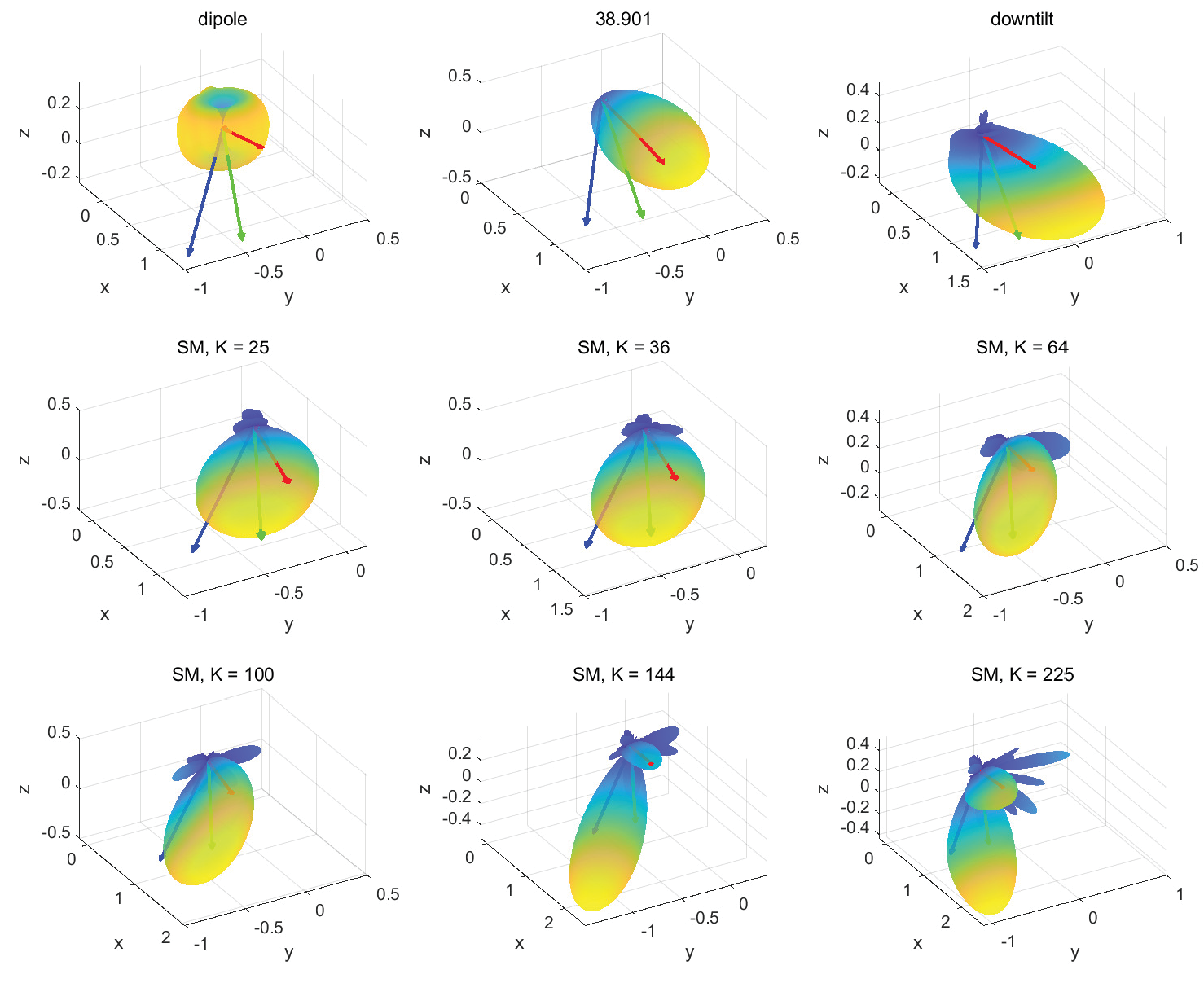}
	\end{center}
	\captionsetup{font={footnotesize}, singlelinecheck=off, name={Fig.}, labelsep=period}
	\caption{Optimized patterns for SM transmission over a LoS-dominated channel to $3$ UEs.}
	\label{Fig.SM_LoS} 
	\vspace*{-5mm}
\end{figure}

In this section, we illustrate the optimized radiation patterns for multi-user scenarios in two typical scenes. 

In Fig. \ref{Fig.SM_LoS}, we consider SM transmission over an LoS-dominant channel (the Rician factor is set to $15$ dB) to $3$ UEs, examining differences in generated radiation patterns for various values of $K$. The optimized radiation patterns are shown in Fig. \ref{Fig.SM_LoS}. The colored arrows denote the angles of departure at the BS towards the different UEs. From Fig. \ref{Fig.SM_LoS}, it is evident that, in this case, the pattern optimization strategy primarily focuses on energy harvesting, aiming to gather as much channel energy as possible from the UEs' main multipath directions. As $K$ increases, the radiation pattern exhibits greater directivity, enabling finer beam steering towards the corresponding users and enhancing the received energy for each path. However, a higher value of $K$ may lead to patterns that are more difficult to realize in hardware, because it is challenging to generate a highly directional pattern with a single antenna. Therefore, in practice, RmMIMO systems with smaller values of $K$ seem more realistic.

\begin{figure}[!t]
	\begin{center}
		\includegraphics[width = 1\columnwidth]{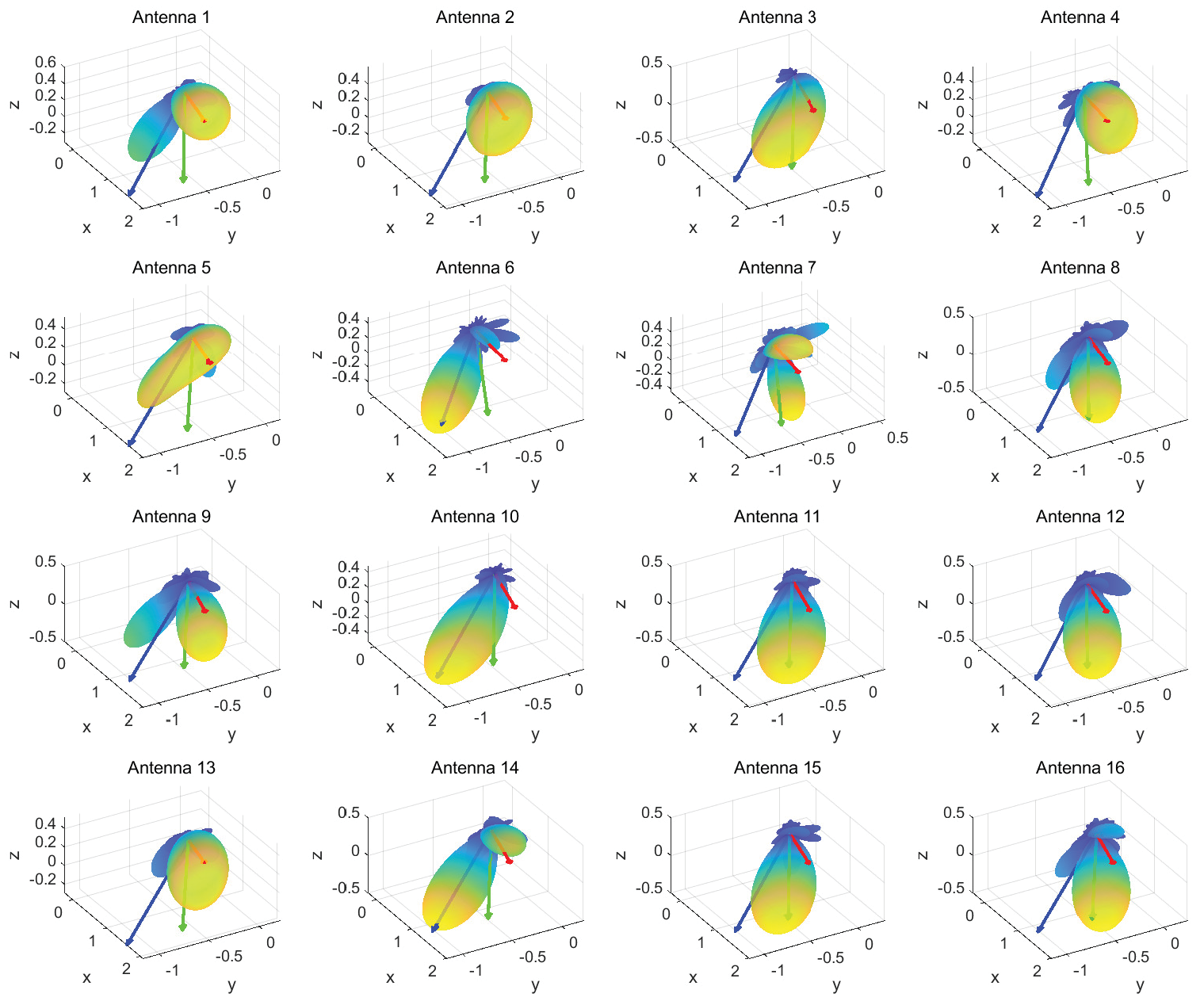}
	\end{center}
	\captionsetup{font={footnotesize}, singlelinecheck=off, name={Fig.}, labelsep=period}
	\caption{Optimized patterns for MM transmission over a LoS-dominated channel to $3$ UEs, where the BS is equipped with a $4\times4$ RmMIMO array.}
	\label{Fig.MM_LoS} 
	\vspace*{-5mm}
\end{figure}

Furthermore, in Fig. \ref{Fig.MM_LoS}, we investigate MM transmission over a LoS-dominated scenario channel, comparing the radiation patterns of different antennas. The number of SH basis functions is set to $K = 64$. It is observed that the main lobe of radiation patterns at different antennas focuses on different paths, thereby harvesting more energy across all directions.

\section{Conclusions}\label{S7}
This paper developed a novel multi-user wideband RmMIMO system that utilizes radiation pattern reconfigurable antennas. By employing the SHOD method, the concept of eCSI was introduced and the antenna radiation pattern design and channel estimation were modeled in the EM domain. The radiation pattern design problem was solved through alternating optimization, and an eCSI estimation scheme was derived using an ESPRIT-based method. Simulation results indicated that customizing radiation patterns to align with the channel characteristics enables RmMIMO to outperform TmMIMO in terms of SE, while an increased design flexibility in radiation patterns leads to larger gains. Furthermore, with the proposed parameterized channel estimation scheme, the RmMIMO system can obtain the necessary eCSI information for radiation pattern design without introducing additional pilot overhead compared to TmMIMO systems. The RmMIMO design shifts from conventional fluid antenna systems, which require antenna physical movement for spatial sampling, to an electronically-controlled system with pixel-based antennas that adjust the radiation pattern, thereby increasing configuration speed and design flexibility. Future research could explore hybrid RmMIMO systems in frequency division duplex, jointly optimize transmitter and receiver patterns, and approximate continuous space radiation pattern optimization results using patterns from discrete space sets.

\begin{appendices}
\section{Comparison between RmMIMO and TmMIMO architectures}\label{apd1}
\begin{figure}[!h]	
	\vspace{-5mm}
	\centering{\includegraphics[width=1\columnwidth,keepaspectratio]{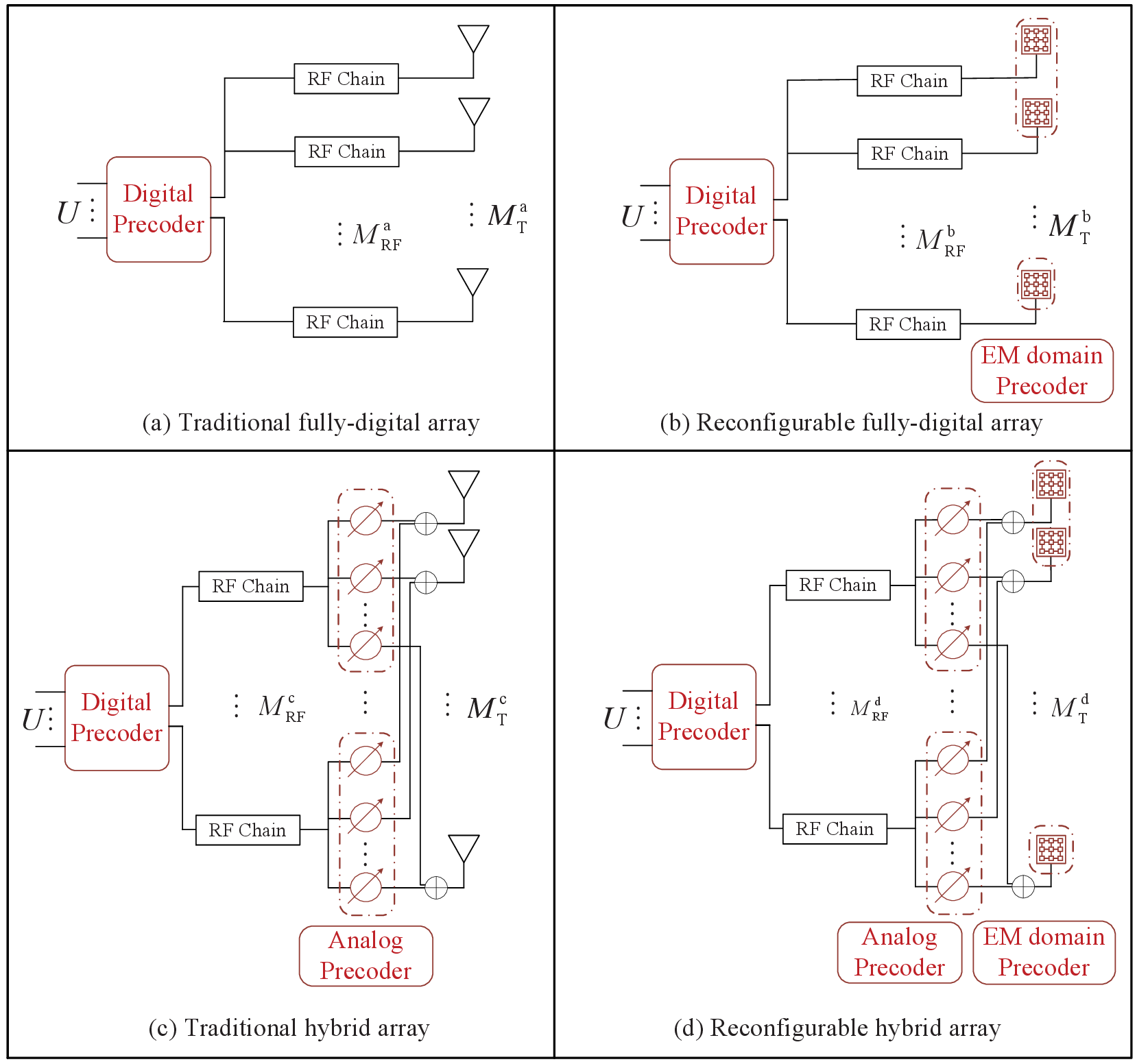}}
	\captionsetup{font={footnotesize}, singlelinecheck=off, name={Fig.}, labelsep=period}
	\caption{Comparison between traditional fully-digital/hybrid arrays and their reconfigurable counterparts.}
	\label{FigI: array_compare}
	\vspace{-3mm}
\end{figure}
Figure \ref{FigI: array_compare} presents a comparison of four different MIMO architectures: (a) Traditional fully-digital array (referred to as TmMIMO in this paper); (b) Reconfigurable fully-digital array (referred to as RmMIMO in this paper); (c) Traditional hybrid array \cite{Heath_HBF}; and (d) Reconfigurable hybrid array.

Figure \ref{FigI: array_compare} reveals that the primary difference between reconfigurable and traditional arrays lies in the antenna configuration. This paper compares architectures (a) and (b) to assess the additional benefits provided by reconfigurable antenna radiation patterns. We refer to the ability to customize radiation patterns as EM domain precoding, distinguishing it from traditional digital and analog precoding. Furthermore, the proposed architecture, although distinct from traditional hybrid arrays (c), is compatible with it. Specifically, by replacing the antennas in the traditional hybrid array with reconfigurable antennas, we obtain structure (d). A detailed comparison of these architectures cannot be provided here due to the limited space. Interested readers may refer to our online supplementary material for more details, see \textit{\url{https://github.com/kekeyingBIT/PRA/blob/main/supplement.pdf}}.

\end{appendices}

\end{document}